\documentclass[twocolumn,showpacs,superscriptaddress,pre]{revtex4}

\usepackage{bm}
\usepackage{amsmath}
\usepackage{amssymb}
\usepackage{amsfonts}
\usepackage{units}
\usepackage{graphicx}
\usepackage[usenames]{color}
\usepackage{subfigure}
\usepackage{lipsum}
\usepackage{soul}

\newcommand{\beq}{\begin{equation}}
\newcommand{\eeq}{\end{equation}}
\newcommand{\bea}{\begin{eqnarray}}
\newcommand{\eea}{\end{eqnarray}}
\newcommand{\bal}{\begin{align}}
\newcommand{\eal}{\end{align}}

\newcommand{\eq}[1]{Eq.~(\ref{#1})}
\newcommand{\eqs}[2]{Eqs.~\eqref{#1}-\eqref{#2}}
\newcommand{\eqa}[2]{Eqs.~\eqref{#1} and \eqref{#2}}

\newcommand{\fig}[1]{Fig.~\ref{#1}}

\newcommand{\rnd}[1]{\!\left(#1\right)}
\newcommand{\sqr}[1]{\!\left[#1\right]}
\newcommand{\abs}[1]{\!\left|#1\right|}

\newcommand{\cly}[1]{\!\left\lbrace#1\right\rbrace}

\newcommand{\dif}{\mathrm{d}}

\newcommand{\kvec}{\mathbf{k}}

\newcommand{\qvec}{\mathbf{q}}
\newcommand{\rvec}{\mathbf{r}}
\newcommand{\Rvec}{\mathbf{R}}

\newcommand{\ofq}{\rnd{\qvec}}
\newcommand{\ofkw}{\rnd{\kvec,\omega}}
\newcommand{\ofkwt}{\rnd{\kvec,\omega; t}}
\newcommand{\ofkwts}{\rnd{\kvec,\omega'; t}}

\newcommand{\retadv}{\text{R/A}}

\newcommand{\ret}{\text{R}}

\newcommand{\adv}{\text{A}}

\begin{document}

\title{Quantum theory for the dynamic structure factor in correlated two-component systems in non-equilibrium -- Application to x-ray scattering}

\author{J. Vorberger}
\email{j.vorberger@hzdr.de}
\affiliation{Institute of Radiation Physics, Helmholtz-Zentrum Dresden-Rossendorf e.V., D-01328 Dresden, Germany}


\author{D.A. Chapman}
\email{david.chapman@firstlightfusion.com}
\affiliation{AWE plc, Aldermaston, Reading RG7 4PR, UK}
\affiliation{Centre for Fusion, Space and Astrophysics, University of Warwick, Coventry CV4 7AL, UK}

\date{\today}


\begin{abstract}
We present a quantum theory for the dynamic structure factors in non-equilibrium, correlated, two-component systems such as plasmas or warm dense matter. The polarization function, which is needed as the input for the calculation of the structure factors, is calculated in non-equilibrium based on a perturbation expansion in the interaction strength. To make our theory applicable for x-ray scattering, a generalized Chihara decomposition for the total electron structure factor in non-equilibrium is derived. Examples are given and the influence of correlations and exchange on the structure and the x-ray scattering spectrum are discussed for a model non-equilibrium distribution, as often encountered during laser heating of materials, as well as for two-temperature systems.
\end{abstract}

\maketitle

\section{Introduction} 

A number of experimental methods exist for the creation and diagnostics of dense states of matter usually only found in massive compact astrophysical objects, but which may routinely be produced in the laboratory for, e.g., inertial confinement fusion experiments \cite{GR_2009,GMMM_2010,FLDG_2015,KDBL_2015}. Typically, the creation of such extreme states involves the rapid deposition of large amounts of energy into the system on time scales ranging from nano- to pico- or even femtoseconds. The properties of these highly transient states generally depend on the duration of the driver, the density of the system and the efficacy of dissipative processes such as radiation, diffusion and equilibration. Apart from fundamental research and laboratory astrophysics, applications in fields such as medical therapy and industrial processes also exist \cite{bauerle_book,BS_2013,VV_2003,FW_2006}.

In all such systems, highly non-equilibrium states are inevitably produced. For instance, the ions may be heated using shock waves \cite{CNXF_1992,NCXF_1995,LNCD_2009} or the electrons may be heated using high-intensity sources of electromagnetic or particle radiation, e.g. optical or x-ray lasers or ion beams. In either case, the species to which the heating mechanism does not couple efficiently are left in the initial state of the undriven system \cite{FSTB_1992,GSL_1995,KBBA_2000,EHHS_2009,WBEV_2016,FBDD_2010,WVBC_2012}. Such systems have been modeled using a variety of numerical techniques including kinetic equations, particle-in-cell simulations and hybrid-fluid models \cite{KBBS_1999,KB_2000,KRVG_2000,RKVG_2002,CG_2011,MZFG_2011,KCDS_2011,MR_2013,MLZ_2015}.

Since non-equilibrium states of matter can be readily created in macroscopic volumes, spatial inhomogeneities and/or anisotropies may occur \cite{KCGR_2016}, but very often isotropic Wigner distribution functions \cite{kremp_book}, strongly deviating from the equilibrium Fermi-Dirac form, dominate the physics \cite{FBDD_2010,CG_2011,LASD_2015}. Once a strongly non-equilibrium state has been produced, the system inexorably relaxes towards full thermal equilibrium. This involves many intricately linked and interesting processes, such as the build up of correlations leading to new structural order, the establishment of well-characterized and distinct electron and ion temperatures through the formation of Fermi-Dirac-shaped distributions within the subsystems, ionization balance, and energy and temperature relaxation between different particle species \cite{SKB_1999,MZFG_2011,SBK_1988,BSP_1998,VCW_2014,VGBS_2010}.

In particular, current experiments combining high-power, short-pulse optical lasers with x-ray free-electron lasers (XFELs), to respectively create and probe warm dense matter (WDM), enable unprecedented insight into the complex microscopic structure of a variety of exotic states \cite{FBDD_2010,FLDG_2015,GRDR_2011,HGDL_2012,ZSHB_2014}. Principally, the  information on the system is contained in the total (bound and free) electron dynamic structure factor, which may be directly measured by spectrally or angularly resolving the  radiation power scattering off the target under study. Such a setup provides an ideal platform for comparing experimental data to theoretical models. By fitting experimental spectra with theoretical calculations, estimates of the plasma conditions, such as the ionization balance, density and mean energy (temperature or Fermi energy for equilibrium systems), may be inferred in addition to the static and dynamic structure or information about collective modes \cite{GLNL_2007,GGGV_2008,KNCD_2008,LNCD_2009,RFGR_2012,MDFF_2013,KVGB_2013,CVFD_2015,FLDG_2015}. X-ray Thomson scattering (XRTS) is therefore envisaged to shed light onto open problems in the understanding of the relaxation of particle momenta and energy, and also temperature equilibration \cite{CNXF_1992,NCXF_1995,WVBC_2012,HBCD_2015,DWP_1998,VGBS_2010}.

While low lying bound state spectra have been investigated in non-equilibrium \cite{HS_2004,VCW_2014}, state of the art theories for the calculation of the total electron structure and scattering spectrum are valid in equilibrium only and can be applied to two-temperature systems only in very limited circumstances \cite{C_2000,GRHG_2007,WVG_2009,FWR_2010,WVGG_2011,VDTG_2012,VGK_2013,VG_2015,PRBS_2015}. Until now, fully non-equilibrium calculations have been restricted to the weakly coupled electron gas \cite{CG_2011, CVFD_2015} or to classical plasmas\cite{RR_1962}. 

In this paper, we present a theoretical model which allows the study of non-equilibrium two-component systems beyond the random phase approximation (RPA). Although we do not consider inhomogeneous or anisotropic systems, we fully account for non-equilibrium Wigner distributions. The evolution of the distribution functions is assumed to be known from other means, such as the solution of kinetic equations or simulations \cite{KBBS_1999,KB_2000}. In order to be able to analyze the scattered signal in non-equilibrium, we generalize the concept of the Chihara decomposition of the total electron structure factor. 

This paper is organized as follows: In Section \ref{sec2}, expressions for the dynamic structure factors for a correlated two-component quantum system in non-equilibrium are derived. Section \ref{sec5} presents the approximation for the polarization function, which includes contributions from vertex and self energy terms. The results for the non-equilibrium and equilibrium structure are presented and discussed in Section \ref{sec6}. In Section \ref{sec3}, a generalized Chihara-like decomposition is derived and expressions for the generalized screening cloud and free electron structure are presented. Finally, examples for such a decomposition of the total structure are shown.


\section{Theoretical description of the dynamic structure factor for non-equilibrium systems\label{sec2}}

The dynamic structure factor (DSF) contains all the information about time-dependent long- and short-range order in interacting many-particle systems. In a non-equilibrium system of particles obeying quantum statistics (such as fermions), the DSF is given by the Fourier transform of the correlation function of density fluctuations $L_{ab}^{>}$ \cite{pines_book,kremp_book}
\begin{align}
 \label{see}
  S_{ab}\ofkwt
  =
  \frac{1}{2\pi n_{ab}}\!
  \int \dif\rvec \dif\tau\,
  e^{-i(\kvec\cdot\rvec-\omega\tau)}
  i\hbar L^{>}_{ab}(12)
  \,.
\end{align}
The labels $a$ and $b$ identify the particle species of interest, with the geometric mean of their mean number densities $n_{ab} = \sqrt{n_{a}n_{b}}$. The microscopic fluctuations of these density fields are given in the position-time basis by $\delta\rho_a(1)= \psi_a^{\dagger}(1)\psi_a(1) -\langle\psi_a^{\dagger}(1)\psi_a(1)\rangle$ with $1=\{\rvec_1,t_1,\sigma_1\}$, where the operators $\psi_a^{\dagger}(1)$ and $\psi_a(1)$ create or annihilate a ket state given by the full set of state variables. We therefore have $i\hbar L_{ab}^{>}(12)=\langle\delta\rho_a(1)\delta\rho_b(2)\rangle$, in which $\langle\ldots\rangle = \text{Tr}\cly{\hat{\varrho}\dots}$ denotes the ensemble average with respect to the non-equilibrium density operator $\hat{\varrho}$ \cite{kremp_book}. In \eq{see}, we have introduced Wigner coordinates related to time and space; the difference coordinates $\tau=t_1-t_2$ and ${\bf r}={\bf r}_1-{\bf r}_2$, and the center-of-mass coordinates $t=\frac{1}{2}(t_1+t_2)$ and ${\bf R}=\frac{1}{2}({\bf r}_1+{\bf r}_2)$. The former broadly represent the scale lengths of microscopic processes, such as density fluctuations, whilst the latter represent macroscopic processes, such as hydrodynamic evolution and spatial gradients. We have suppressed the macroscopic space variable $\Rvec$ as we consider homogeneous systems only.

Equation \eqref{see} represents a general form of the fluctuation-dissipation theorem. This is usually understood in the context of equilibrium systems, wherein the density fluctuations are directly connected to the imaginary (dissipative) part of the retarded density response of the system to the applied field. In order to provide a valid description for non-equilibrium systems, we require a suitably general framework such as that provided by the Keldysh formalism using real time non-equilibrium Green's functions \cite{K_1965}. The equation of motion for $L_{ab}$ defined on the Keldysh contour ${\cal C}$ is given by \cite{kremp_book}
\begin{flalign}
\label{L_ab_eqn_of_motion}
L_{ab}(12) 
  =&\,
  \Pi_{ab}(12)
  \nonumber\\
  & + \sum_{c,d}\! \int\limits_{\cal C}\! \dif 3\dif 4 \, \Pi_{ac}(13)  V_{cd}(34) L_{db}(42)\,.
\end{flalign}
Here, $\Pi_{ab}$ is the polarization function, which determines the response of the density field of species $a$ to changes in the effective field in the system due to species $b$. For fermions, one has $\Pi_{ab}(12,1'2')= -i\hbar\,\delta g_a(11')/\delta U_b^{\text{eff}}(2'2)$ \cite{kremp_book}. The two-point function required in \eq{L_ab_eqn_of_motion} is given by $\Pi_{ab}(12) = \Pi_{ab}(12,1'\!\to\!1^{+},2'\!\to\!2^{+})$, where $1^{+} = \cly{\rvec_{1},t_{1}^{+},\sigma_{1}}$ represents an event at point $\rvec_{1}$ at an infinitesimally later time than $t_{1}$. The resulting time ordering is crucial for non-equilibrium systems, which do not obey the adiabatic theorem. Interactions between the particles are mediated by the unscreened Coulomb interaction $V_{ab}(12) = V_{ab}\rnd{\rvec_{1} - \rvec_{2}}\delta\rnd{t_{1} - t_{2}}$. Thus, dynamic srceening is entirely determined by the response functions of the system and, in particular, the polarization functions.

Equations for the correlation functions $L^{\gtrless}_{ab}$ and the retarded and advanced functions $L^{\retadv}_{ab}$ can be obtained from \eq{L_ab_eqn_of_motion} using the Keldysh techniques \cite{K_1965}. In contrast to equilibrium systems, the correlation functions are needed in addition to retarded and advanced quantities. This is due to the fact that the Kubo-Martin-Schwinger relation does not hold in non-equilibrium \cite{kremp_book}.

\subsection{Density response of fully interacting two-component systems}

\allowdisplaybreaks
For a system containing an arbitrary number of particle species, the DSF can easily be represented using a very general matrix notation. We limit the present discussion to the important case of two-component (e.g. electron-ion) systems as the roles of interactions between the species are more clearly elucidated and the structure of the response functions can be solved for analytically. Following the same general route as used in Ref.~\cite{VGBS_2010}, we rewrite the system of equations generated by \eq{L_ab_eqn_of_motion} as
\newcommand{\lcal}{{\cal L}}
\newcommand{\rcal}{{\cal R}}
\newcommand{\qcal}{{\cal Q}}
\newcommand{\tcal}{{\cal T}}
\begin{align}
  \label{L_ab_def}
  L_{ee}
  = &\,
  \lcal_{ee} + \rcal_{ee} + (\rcal_{ee}V_{ee} + \rcal_{ei}V_{ie})L_{ee}
  \,,
  \nonumber\\
  L_{ei}
  = &\,
  \lcal_{ei} + \rcal_{ei} + (\rcal_{ei}V_{ie} + \rcal_{ee}V_{ee})L_{ei}
  \,.
\end{align}
Equivalent expressions for the functions $L_{ii}$ and $L_{ie}$ are generated by interchanging the species labels $e\Leftrightarrow i$ in every term in \eq{L_ab_def}. Here, we have omitted writing both the dependencies on the spatio-temporal variables and also the integrations for brevity. Each of these equations should be read to have a structure identical to \eq{L_ab_eqn_of_motion}. In \eq{L_ab_def}, we have introduced the following auxiliary response functions which collect certain subsystem contributions
\begin{align}
  \label{lcal_ab_def}
  \lcal_{ee}
  = &\,
  \Pi_{ee} + (\Pi_{ee}V_{ee} + \Pi_{ei}V_{ie}) \lcal_{ee}\,, 
  \nonumber\\
  \lcal_{ei}
  = &\,
  \Pi_{ei} + (\Pi_{ee}V_{ee} + \Pi_{ei}V_{ie})\lcal_{ei}
  \,,
 \end{align}
 and
 \begin{align}
  \label{rcal_ab_def}
  \rcal_{ee}
  = &\,
  \lcal_{ee}V_{ei}\lcal_{ie} + \lcal_{ei}V_{ii}\lcal_{ie}\,,
  \nonumber\\
  \rcal_{ei}
  = &\,
  \lcal_{ee}V_{ei}\lcal_{ii} + \lcal_{ei}V_{ii}\lcal_{ii}\,. 
\end{align}
Again, the structure of the set of \eqs{lcal_ab_def}{rcal_ab_def} in the space-time domain has the form of \eq{L_ab_eqn_of_motion} and the expressions for the ion-ion an ion-electron functions are found by interchanging of species labels.

Upon transferring from the Keldysh contour to the physical time axis, the Langreth-Wilkins rules \cite{LW_1972} are used to obtain the correlation functions and corresponding retarded/advanced functions. For the electron-electron density fluctuation correlation function required in \eq{see}, one finds
\begin{widetext}
\begin{align}
  \label{L_ee_>_def}
  L_{ee}^{>}(\rvec_{1}\rvec_{2},t_{1}t_{2})
  = &\,
  \lcal_{ee}^{>}(\rvec_{1}\rvec_{2},t_{1}t_{2}) + \rcal_{ee}^{>}(\rvec_{1}\rvec_{2},t_{1}t_{2})
  \nonumber\\
  & 
  + \int \dif\rvec_{3} \dif \rvec_{4}
  \int\limits_{-\infty}^{+\infty} \dif t_3
  \left\{
  \vphantom{\frac{1}{1}}\!
  \left[\rcal^{>}_{ee}(\rvec_{1}\rvec_{3},t_1t_3)\,V_{ee}(\rvec_{3} - \rvec_{4})
  + \rcal^{>}_{ei}(\rvec_{1}\rvec_{3},t_1t_3)\,V_{ie}(\rvec_{3} - \rvec_{4}) \right] L^{\adv}_{ee}(\rvec_{4}\rvec_{2},t_3t_2)
  \right.
  \nonumber\\
  &+ \left. 
  \left[\rcal^{\ret}_{ee}(\rvec_{1}\rvec_{3},t_1t_3)\,V_{ee}(\rvec_{3} - \rvec_{4})
  + \rcal^{\ret}_{ei}(\rvec_{1}\rvec_{3},t_1t_3)\,V_{ie}(\rvec_{3} - \rvec_{4})\right] 
  L^{>}_{ee}(\rvec_{4}\rvec_{2},t_3t_2)
  \vphantom{\frac{1}{1}}
  \right\}\,.
\end{align}
\end{widetext}
All the dependencies on the spatial and temporal coordinates have been restored for clarity. Note that the integration over $t_{4}$ is eliminated by the fact that the Coulomb potential is represented as being local in time. Similar expressions for the ion-ion, electron-ion and ion-electron density fluctuation correlation functions can also be obtained and are detailed in Appendix A.

We now briefly discuss the importance of the Wigner coordinates for non-equilibrium systems. Writing the internal coordinates of the integrations in \eq{L_ee_>_def} in terms of new difference and center-of-mass coordinates, the micro- and macroscopic scales become inextricably coupled which prevents straightforward Fourier transformation of $L_{ee}^{>}(12)$. Performing a gradient expansion with respect to the internal difference coordinates, the macroscopic spatio-temporal scales enter only parametrically at lowest order (the local approximation) \cite{kremp_book}. Higher order corrections connected with evolving spatially inhomogeneous systems have recently been considered \cite{KCGR_2016}. For the present work, we are motivated by experiments which probe the high-frequency (short time scale) response of small and relatively homogeneously heated systems and, thus a local approximation is sufficient. In this case, equation \eqref{L_ee_>_def} yields convolution-like structures and is algebraic in Fourier space. The framework presented in this work is reasonable for incorporating spatial inhomogeneity for shallow gradients.

Based on the full set of results for the various correlation and retarded/advanced functions (Appendix A), the electronic density fluctuation correlation function of a fully interacting two-component system in frequency-momentum space is found to be
\begin{align}
  \label{L_ee_>_result}
  L_{ee}^{>}
  = &\,
  \frac{(1 - \tcal_{ee}^{\adv})\qcal_{ee}^{>} + \tcal_{ee}^{>}\qcal_{ee}^{\adv}}{|1 - \tcal_{ee}^{\ret}|^2}
  \,,
\end{align}
where
\begin{align}
  \label{Q_ee_X_def}
  \qcal_{ee}^{\text{X}}
  = &\,
  \lcal_{ee}^{\text{X}} + \rcal_{ee}^{\text{X}}
  \,,
  \\
  \label{T_ee_X_def}
  \tcal_{ee}^{\text{X}}
  = &\,
  \rcal_{ee}^{\text{X}}V_{ee} + \rcal_{ei}^{\text{X}}V_{ie}
  \,.
\end{align}
with the label $\text{X} \to \,\gtrless$ or $\retadv$, as required. In \eqs{L_ee_>_result}{T_ee_X_def}, all functions now depend on the set of Fourier variables $\cly{\kvec,\omega}$ and are parametrized by the macroscopic time $t$, e.g.~$L_{ee}^{>} \equiv L_{ee}^{>}\ofkwt$. The exception is the Coulomb potential, which depends only on the wave number, i.e.~$V_{ab} \equiv V_{ab}(k)$. The other correlation functions of interest for the ion-ion, electron-ion and ion-electron structure factors are given in Appendix A.

In \eq{L_ee_>_result}, the denominator term $|1-\tcal_{ee}^{\ret}|^{2}$ acts as a generalization of the two-component dielectric function, which gives the location of all collective excitations (quasi-particles and single-particle modes) and describes dynamic screening, exchange and correlations. The numerator term can be interpreted as the spectral function of the system, which describes the occupations of the possible states (excitations) and their lifetimes. The complex structure of \eq{L_ee_>_result} in terms of $\lcal_{ab}$ is the result of the interplay of correlations both within and between the distinct electron and ion subsystems. In particular, note that pure electron or pure ion subsystem correlations cannot be separated out due to contributions from the cross species terms $\Pi_{ei}$ and $\Pi_{ie}$.

Finally, the correlation and retarded/advanced functions for the auxiliary quantities $\lcal_{ab}$ must be considered. These are found to be
\begin{align}
  \label{lcal_ee_ei_<>}
  \lcal_{ee}^{>}
  =&\,
  \frac{(1 - \Pi_{ei}^{\adv}V_{ie})\,\Pi_{ee}^{>} + \Pi_{ei}^{>}V_{ie}\Pi_{ee}^{\adv}}
  {|1 - \Pi_{ee}^{\ret}V_{ee} - \Pi_{ei}^{\ret}V_{ie}|^2}
  \,,
  \nonumber\\
  \lcal_{ei}^{>}
  =&\,
  \frac{(1 - \Pi_{ee}^{\adv}V_{ee})\,\Pi_{ei}^{>} + \Pi_{ee}^{>}V_{ee}\Pi_{ei}^{\adv}}
  {|1 - \Pi_{ee}^{\ret}V_{ee} - \Pi_{ei}^{\ret}V_{ie}|^2}
  \,,
  \\
  \label{lcal_ee_ei_R/A}
  \lcal_{ee}^{\retadv}
  =&\,
  \frac{\Pi_{ee}^{\retadv}}{1 - \Pi_{ee}^{\retadv}V_{ee} - \Pi_{ei}^{\retadv}V_{ie}}
  \,,
  \nonumber\\
  \lcal_{ei}^{\retadv}
  =&\,
  \frac{\Pi_{ei}^{\retadv}}{1 - \Pi_{ee}^{\retadv}V_{ee} - \Pi_{ei}^{\retadv}V_{ie}}
  \,.
\end{align}
Substituting these results into \eq{L_ee_>_result}, it is clear that the quantities of principal importance for providing a theoretical basis for the DSF are the correlation functions (and corresponding retarded/advanced functions) for the irreducible polarisabilities $\Pi_{ab}$. Appropriate expressions for the latter are the focus of the next section. 

\subsection{Diagonalised polarization approximation - linear response}

If only diagonal elements to the polarization function are considered, i.e. $\Pi_{ei} = \Pi_{ie} = 0$, a significant simplification to the fully interacting density response \eqref{L_ee_>_result} can be made that enables a more tractable treatment of many related phenomena. Under this diagonalized polarization approximation (DPA), one has $\lcal_{ei} = \lcal_{ie} = 0$ and $\rcal_{ee} = \rcal_{ii} = 0$ and \eq{L_ee_>_result} becomes
\begin{align}
\label{Lee_no_pi_ei}
  L_{ee}^{>}
  = &\,
  \frac{\lcal_{ee}^{>} + |\lcal_{ee}^{\ret}\!|^2V_{ei}^2\lcal_{ii}^{>}}
  {|1-V_{ie}\lcal_{ee}^{\ret} V_{ei}\lcal_{ii}^{\ret}|^{2}}
  \,.
\end{align}
The physical interpretation of the DPA is that the average one-particle state of a particle of species $a$ does not depend on the effective external potential due to species $b$. This does not, however, imply that all correlations between species $a$ and $b$ are formally neglected; the self energy of species $a$ may still contain inter-species correlations. However, the {\em direct} electron-ion coupling is now implied to be weak and a two fluid description may be used. An important example for such a system is metallic hydrogen, which can be treated as a strongly coupled proton fluid coexisting with a highly degenerate electron gas, i.e. a Lorentz plasma \cite{KDBL_2015}. Since strong coupling within independent subsystems may still be included in DPA, we remark that a clear distinction to the well-known random phase approximation (RPA), wherein {\em all} inter-particle interactions are taken to be weak, must be made, despite the resulting structure and response functions being identical in form.

The first term in the numerator of Eq.~\eqref{Lee_no_pi_ei} is the pure electron gas contribution \cite{CG_2011}. It is followed by the ionic structure contributions to the total electronic structure (a convolution of the electronic screening cloud with the bare ion structure). The two subsystems are coupled via the denominator. In the case of the DPA, as presented here, this coupling is given in linear response and mediated by average fields. A consequence of this coupling is given by, e.g., the screening of the ionic plasmon mode that produces ion acoustic waves \cite{VG_2009,VGBS_2010}. 

In DPA, the auxiliary functions $\lcal_{aa}$ can be interpretted as the density response functions of independent electron and ion subsystems, which are coupled by Coulomb interactions through the functions $\rcal_{ab}$. In this case, the DSF can then be tractably expressed in terms of the polarization functions only
\begin{align}
 \label{See_DPA}
 S_{ee}\ofkwt
 = &\,
 \frac{\abs{1 - \Pi_{ii}^{\ret}\ofkwt V_{ii}(k)}^{2}}{\abs{\varepsilon\ofkwt}^{2}}\frac{i\hbar\Pi_{ee}^{>}\ofkwt}{2\pi n_{e}}
  \nonumber \\
 & + 
 \frac{\abs{\Pi_{ee}^{\ret}\ofkwt V_{ei}(k)}^{2}}{\abs{\varepsilon\ofkwt}^{2}}\frac{i\hbar\Pi_{ii}^{>}\ofkwt}{2\pi n_{i}},
\end{align}
where the retarded dielectric screening function of the system has the usual form $\varepsilon = 1 - \sum_{a}\Pi_{aa}^{\ret} V_{aa}$. Expressions of the general form of \eq{See_DPA} were derived by Rosenbluth and Rostocker \cite{RR_1962}, although only for non-degenerate systems in RPA.


\section{Polarization functions of non-equilibrium systems\label{sec5}}

\begin{figure}[t]
  \includegraphics[width=0.40\textwidth,clip=true]{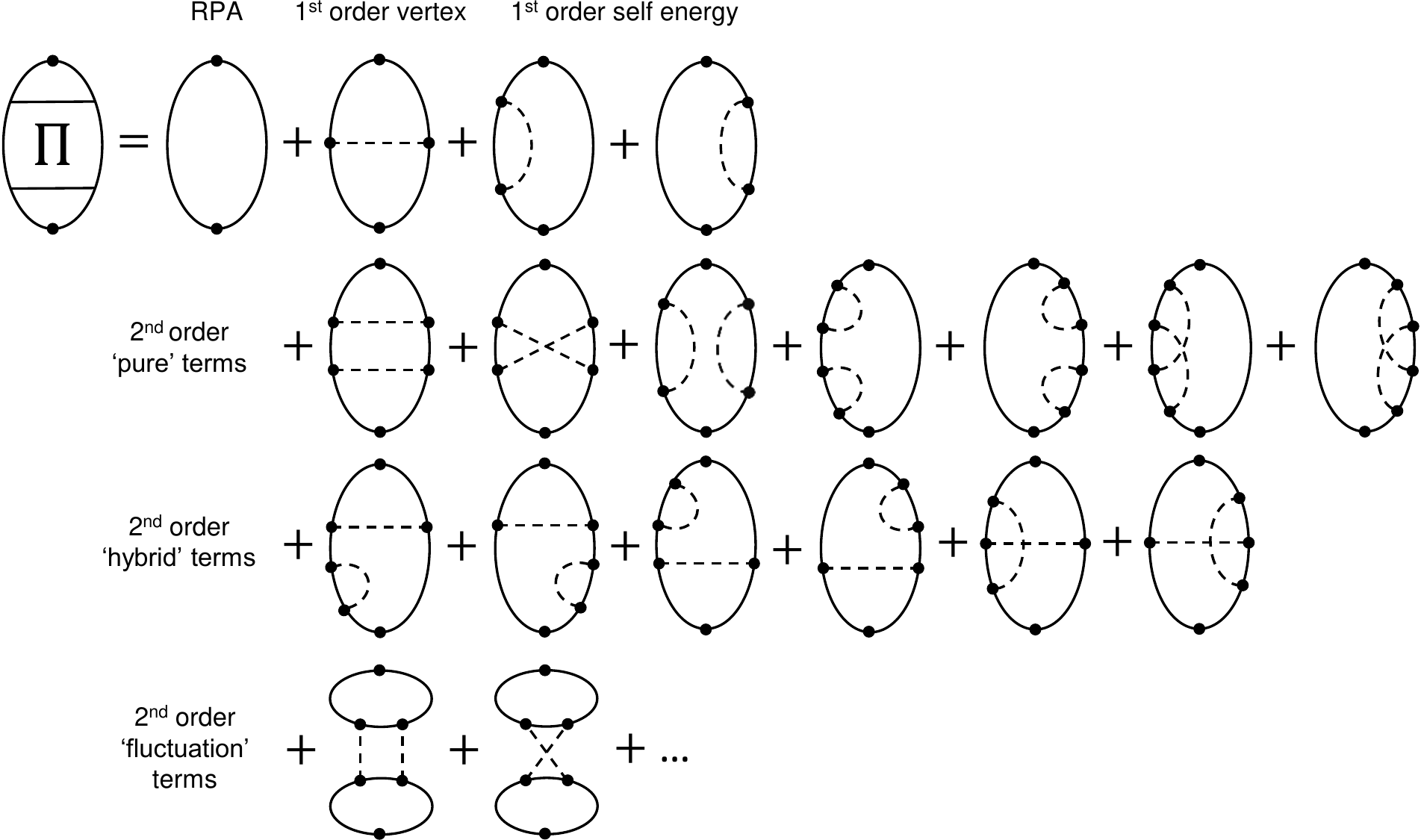}
  \caption{
   Diagrammatic representation of the expansion of the Bethe-Salpeter equation for the polarization function $\Pi$. Terms up to second order in the screened interaction $V_{aa}^{\text{sc}}$ (dashed lines) are shown. 
    }
    \label{pi_expansion}
\end{figure}

The equation of motion for the polarization function obeys a Bethe-Salpeter equation depending on the self-consistent single-particle Green's functions and screened self energy \cite{kremp_book}. This very general equation of motion cannot presently be numerically solved for arbitrary coupling strengths due to the presence of functional derivatives. Instead, it is possible to establish a perturbation expansion with respect to the interaction strength (see \fig{pi_expansion}), which leads to corrections to the RPA. In this work, we retain the exact first order corrections accounting for self energy and exchange. Such a method has been used before to various perturbation orders, even including the full vertex function \cite{GT_1970a,GT_1970b,HAS_1979,GNS_1985,CS_1988,EV_1990,FM_1992,RA_1994,HA_1997}. However, all these previous calculations have been performed for equilibrium systems, which greatly simplifies matters since only retarded quantities are needed. Furthermore, as we are motivated by current experiments producing and studying warm dense matter, we need dynamic corrections to the RPA covering a broad range of states including both degenerate and non-degenerate systems.

At first order in the interaction, there are two terms in addition to the RPA \cite{kremp_book,VSK_2004}
\begin{align}
  \label{pi_series}
  \Pi_{ab}^{1}(12) 
  = 
  \delta_{ab} \rnd{\Pi_a^{0}(12) + \Pi_a^{\text{V}}(12) + \Pi_a^{\text{S}}(12)}\,,
\end{align}
where $\Pi_a^{0}$ is the RPA contribution. The additional first order terms are identified as a {\em vertex} correction $\Pi_a^{\text{V}}$ and a {\em self energy} correction $\Pi_a^{\text{S}}$. 

As shown in \fig{pi_expansion}, fifteen terms exist at second order. The first two of these result from iteration of the screened ladder and are therefore pure second order vertex contributions, the next five correspond to pure second order self energy contributions, and a further six represent hybrid terms containing both vertex and self energy characteristics. All of these additional terms contain $\delta_{ab}$ and, thus, are all single species terms similar to \eq{pi_series}. The final two second order terms are the {\em fluctuation} contributions and are the first to give non-vanishing contributions to $\Pi_{ab}$ with $a\neq b$. The dominance of single-species terms at this level of approximation may explain why the two fluid approach, which neglects direct electron-ion polarization, is so successful in describing many systems. Furthermore, as cross-species terms are all of second order or higher, it is unlikely that a perturbative treatment of these will be sufficient and a full summation of the ladder will be required once {\em direct} electron-ion correlations need to be included.
 

\allowdisplaybreaks
\subsection{Random phase approximation}
The RPA term has been considered in non-equilibrium before \cite{CG_2011}. We give the correlation functions of the RPA contribution here for completeness
\begin{align}
  \label{rpa_><}
  {\Pi_{a}^{0}}^{\gtrless}(12) 
  = 
  -i\hbar s_{a} {g_{a}^{0}}^{\gtrless}(12) \, {g_{a}^{0}}^{\lessgtr}(21)
  \,,
\end{align}
where $s_{a} = 2\sigma_{a} + 1$ gives the summation over the particle spin for fermions. The free single particle correlation functions are averages over creation and annihilation operators $ig^<(12)=\pm\langle\psi^{\dagger}(2)\psi(1)\rangle$ and $ig^>(12)=\pm\langle\psi(1)\psi^{\dagger}(2)\rangle$. Fourier transformation of Eq. \eqref{rpa_><} with respect to the microscopic variables $\rvec$ and $\tau$ proceeds straightforwardly in local approximation, resulting in a convolution of the single particle correlation functions with respect to both the wave number $\kvec$ and frequency $\omega$. The correlation functions are given by the spectral function $a_a\ofkwt$ and the Wigner distributions
\begin{align}
  \label{gs}
  g_{a}^{\gtrless}\ofkwt
  = &\,
  i\,a_a\ofkwt f_a^{\gtrless}(\hbar\omega;t)
  \,,
  \\
  \label{gs_2}
  f_a^{\gtrless}(\hbar\omega;t)
  = &\,
  \left\{ 
  \begin{array}{ll}
    f_{a}(\hbar\omega;t) &\quad\gtrless \rightarrow <
    \\
    -[1-f_{a}(\hbar\omega;t)]&\quad\gtrless \rightarrow >
  \end{array}
  \right.
  \,.
\end{align}
The spectral function is provided by the Kadanoff-Baym ansatz \cite{kremp_book}
\begin{align}
  \label{gs_1}
  a_a\ofkwt = &\,
  2\pi i\,\delta(\hbar\omega - E_{a}(\kvec;t))
  \,,
\end{align}
where $E_{a}(\kvec;t) = E_{a}^{0}(\kvec) + \text{Re}\Sigma_{a}(\kvec;t)$, $E_{a}^{0}(\kvec) = \hbar^{2}k^{2}/2m_{a}$ is the kinetic energy of free (non-interacting) particles and $\Sigma_{a}(\kvec;t)$ is the static self energy. The expression \eqref{gs_1} amounts to a relatively simple approximation since it does not account for the finite lifetime (damping) of the excitations and is therefore restricted to the quasi-particle picture. Improvements to the Kadanoff-Baym ansatz (see, e.g., \cite{LSV_1986,F_2009}) are beyond the scope of the present work.

In RPA, only free-particle dispersion relations are considered, i.e.~$E_{a}(\kvec) = E_{a}^{0}(\kvec)$. For convenience we therefore introduce the notation $f_a^{\gtrless}(\kvec;t) \equiv f_a^{\gtrless}(E_{a}^{0}(\kvec);t)$. The Fourier transform of \eq{rpa_><} follows as \cite{kremp_book,CG_2011}
\begin{align}
  {\Pi_a^{0}}^{\gtrless}\ofkwt
  = &\, 
  2\pi i \int\!\frac{\dif\qvec}{(2\pi)^3}\,
  f_a^{\gtrless}(\qvec+\kvec;t)
  f_a^{\lessgtr}(\qvec;t)
  \nonumber\\
  & \times 
  \delta\big( \hbar\omega - \Delta E_{a}^{0}(\qvec,\kvec) \big)
  \,.
\end{align}
where $\Delta E_{a}^{0}(\qvec,\kvec) = E_{a}^{0}(\qvec + \kvec) - E_{a}^{0}(\qvec)$ is the change in kinetic energy due to the momentum shift $\hbar\kvec$. Once the correlation functions are known, the retarded quantities may be obtained via the Kramers-Kronig relation \cite{kremp_book}
\begin{align}
\label{pir}
  \Pi^{\ret}\ofkwt
  =&\,
  -\!\int\limits_{-\infty}^{+\infty}\frac{\dif\omega'}{\pi}\,
  \frac{\text{Im}\Pi^{\ret}\ofkwts}{\omega+i\epsilon-\omega'}
  \,.
\end{align}
with $\text{Im}\Pi^{\ret} = \frac{i}{2}\rnd{\Pi^{<} - \Pi^{>}}$ and $\epsilon \to 0^{+}$. Note that Eq.~\eqref{pir} holds for all combinations of particle labels, including the cross terms. The explicit result for the RPA case is
\begin{align}
  \label{pi_rpa}
  {\Pi_{a}^{0}}^{\ret}\ofkwt
  =
  \int\!\frac{\dif\qvec}{(2\pi)^3}
  \frac{f_a(\qvec;t)-f_a(\qvec+\kvec;t)}
  {\hbar\omega + i\epsilon - \Delta E_{a}^{0}(\qvec,\kvec)}\,.
\end{align}
The complex expressions \eqref{pir} and \eqref{pi_rpa} can be split into their constituent real and imaginary parts using the well-known Dirac-Plemlj identity \cite{kremp_book}.


\subsection{First order vertex correction}
The first term beyond the RPA contribution in \eq{pi_series} is the vertex correction due to interactions between screening particles. In the space and time domain one finds \cite{DSSK_1995}
\begin{align}
 \label{pi_vertex_def}
 \Pi_{a}^{\text{V}}(12)
 = &\,
 -(i\hbar)^{2}s_{a}\int_{{\cal C}} \dif3\dif4\,{g_{a}^{0}}(13){g_{a}^{0}}(32)
 \nonumber\\
 & \times
 V_{aa}^{\text{sc}}(34) {g_{a}^{0}}(24) {g_{a}^{0}}(41)
 \,.
\end{align}
The time integration is over the Keldysh contour. To lowest order, the dynamically screened interaction potential can be taken to be local in time, i.e.~$V_{aa}^{\text{sc}}(34) = V_{aa}^{\text{sc}}(\rvec_3-\rvec_4;t)\delta(t_3-t_4)$. We have retained a macroscopic time dependence in the latter to account for any evolution in the plasma conditions that affect screening. 

The convergence of the vertex term \eqref{pi_vertex_def} does not depend on screening (a bare Coulomb potential is also sufficient) since the term describes exchange and is therefore naturally restricted to short ranges. Nevertheless, a screened potential seems more appropriate to use in systems such as WDM, where screening is known to be important (see, e.g.~Ref.~\cite{CVFD_2015}). The vertex term is responsible for the appearance of the normal $e^4$-exchange term in equation of state theory \cite{VSK_2004}. 

Using the locality of the potential in time, we write
\begin{align}
  \label{piv_keldysh}
  \Pi_a^{\text{V}}(12)
  = &\,
  -(i\hbar)^2 s_{a} \int \dif\rvec_3 \dif\rvec_4\, V_{aa}^{\text{sc}}(\rvec_3-\rvec_4;t)
  \nonumber\\
  & \times
  \int\limits_{\cal C}\! \dif t_3\,
  {\cal G}_{13,41}(t_1,t_3){\cal G}_{32,24}(t_3,t_2)\,,
\end{align}
with ${\cal G}_{13,41}(t_1,t_3)={g_{a}^{0}}(\rvec_1\rvec_3,t_{1}t_3){g_{a}^{0}}(\rvec_4\rvec_{1},t_3 t_1)$ and similarly for ${\cal G}_{32,24}(t_3,t_2)$. The transition from the Keldysh contour onto the physical time axis is again performed using the Langreth-Wilkins rules. Fourier transformation of \eq{piv_keldysh} proceeds with the definition of new internal Wigner coordinates. All the macroscopic contributions are dropped to enforce the local approximation. One finds for the correlation functions
\begin{align}
  \label{piv<>}
  {\Pi_a^{\text{V}}}^{\gtrless}\ofkwt
  =&\,
  2\pi i s_{a} \int \! \frac{\dif\qvec}{(2\pi)^3}\,
  f_a^{\gtrless}(\qvec + \kvec;t)f_a^{\lessgtr}(\qvec;t)
  \nonumber\\ &
  \times
  \delta\big( \hbar\omega - \Delta E_{a}^{0}(\qvec,\kvec) \big)
  \nonumber\\ &
  \times
  I_{a}^{\text{V}}(\qvec,\kvec,\omega;t)
  \,,
\end{align}
where we have defined a vertex correction function
\begin{align}
  \label{piv<>_inv_D}
  I_{a}^{\text{V}}(\qvec,\kvec,\omega;t)
  = &\,
  2\,{\cal P}\! \int \! \frac{\dif\qvec'}{(2\pi)^3}
  V_{aa}^{\text{sc}}(\qvec - \qvec';t)
  \nonumber \\ &
  \times
  \frac{f_a(\qvec';t) - f_a(\qvec'+\kvec;t)}
  {\hbar\omega - \Delta E_{a}^{0}(\qvec',\kvec)}\,.
\end{align}
Here, $\cal P$ denotes a Cauchy principal value integration. In order to obtain \eq{piv<>}, we have used the free single-particle correlation functions ${g_{a}^{0}}^{\gtrless}\ofkw$ as prescribed by Eqs.~\eqref{gs}-\eqref{gs_1}.

The retarded function related to the vertex correction follows from \eq{pir}
\begin{align}
  \label{pirwitt}
  {\Pi_a^{\text{V}}}^{\ret}\ofkwt
  = &\,
  -s_{a} \int\frac{\dif\qvec\,\dif\qvec'}{(2\pi)^6}
  V_{aa}^{\text{sc}}(\qvec-\qvec';t)
  \nonumber \\ &
  \times
  \frac{f_a(\qvec;t)-f_a(\qvec+\kvec;t)}
  {\hbar\omega + i\epsilon - \Delta E_{a}^{0}(\qvec,\kvec) }
  \nonumber \\ &
  \times
  \frac{f_a(\qvec';t)-f_a(\qvec'+\kvec;t)}
  {\hbar\omega + i\epsilon - \Delta E_{a}^{0}(\qvec',\kvec) }\,.
\end{align}
An expression identical in form to \eq{pirwitt} was obtained for the specific case of thermal equilibrium by De Witt {\em et al.}~using the imaginary time Matsubara technique \cite{DSSK_1995}.

For the most general case, where the momentum distributions are anisotropic \cite{WP_1979,YHK_2011}, numerical evaluation of \eqref{piv<>} is computationally expensive. Fortunately, in dense systems such as WDM the collision rate between electrons and ions/neutral atoms is often sufficient to randomize the particle momenta whilst not significantly altering their kinetic energies. Thus, an isotropic approximation is sufficient for our purpose. It is convenient to use a bi-spherical coordinate system in which the wave vector $\kvec$ is defined to be aligned with the positive $z$ axis. The degree of coupling between the dummy integration vectors $\qvec$ and $\qvec'$ in \eqref{piv<>} is entirely determined by the form of the interaction $V_{aa}^{\text{sc}}$. For a simple screened potential such as the Debye interaction one has
\begin{align}
  \label{debye_potential}
  V_{aa}^{\text{sc}}(\qvec;t)
  =
  \frac{Z_{a}^{2}e^{2} 4\pi k_{\text{C}}}{q^{2} + \kappa_{e}^{2}(t)}\,,
\end{align}
in which $Z_{a}e$ is the charge of particles of species $a$, $k_{\text{C}} = 1/4\pi\varepsilon_{0}$ is the Coulomb constant and $\kappa_{e}(t)$ is the (time-dependent) inverse screening length of the electrons. Under isotropic, non-equilibrium conditions, the latter can be written \cite{GVWG_2010}
\begin{align}
  \label{inverse_screening_length}
  \kappa_{e}^{2}(t)
  =
  \frac{4}{\pi a_{\text{B}}}
  \int\limits_{0}^{\infty}\dif q \, 
  f_{e}(q;t)
  \,,
\end{align}
with the Bohr radius $a_{\text{B}} =\hbar^2/m_e e^2 k_C \approx 0.529\,\unit{\AA}$. 

Using \eqa{debye_potential}{inverse_screening_length}, we are able to perform three of the six integrations in Eq.~\eqref{piv<>} analytically. The result is
\begin{align}
  \label{piv<>_numerical}
  {\Pi_{a}^{\text{V}}}^{\gtrless}(k,\omega;t)
  = &~
  \frac{im_{a}s_{a}}{2\pi \hbar^{2} k}\!
  \int\limits_{q_{\min}^{-}}^{\infty} \dif q \, q \,
  f_a^{\gtrless}(E_{a}(q) + \hbar\omega;t) 
  \nonumber\\ &
  \times 
  f_a^{\lessgtr}(q;t) 
  I_{a}^{\text{V}}(q,k,\omega;t)\,,
  \\
  \label{piv<>_numerical_2}
  I_{a}^{\text{V}}(q,k,\omega;t)
  = &~
  \frac{2Z_{a}^{2}m_{a}}{\pi k a_{\text{B}}m_{e}}
  \frac{1}{q}
  \int\limits_0^{\infty} \dif q' \, {\cal P} \! \int\limits_{-1}^{1} \frac{\dif\zeta}{\zeta - \zeta_{0}} \, 
  \nonumber\\ &
  \times  
  \frac{f_a(q';t) - f_a(\tilde{q}';t)}{\sqrt{\abs{a^{2} - b^{2}}}}\,.  
\end{align}
The shifted dummy wave number appearing in the second line of \eq{piv<>_numerical_2} is given by $\tilde{q}' = \sqrt{(q')^{2} + k^{2} - 2 q'k \zeta}$. The summands $a$ and $b$ in the square root term are functions of the remaining integration variables
\begin{align}
  a
  = &\,
  \frac{1}{2}\rnd{\frac{q}{q'} + \frac{q'}{q} + \frac{\kappa_{e}^{2}}{qq'}} - \zeta \xi_{0} \,,
  \\
  b
  = &\,
  \sqrt{(1 - \zeta^2)(1 - \xi_{0}^2)}\,.
\end{align}
We also have the special values
\begin{align}
  \xi_{0}
  = &\,
  \frac{m_a}{\hbar^{2}kq}\!\left(\frac{\hbar^{2}k^{2}}{2m_a} - \hbar\omega\right)\,,
  \\
  \zeta_{0} 
  = &\,
  \frac{m_a}{\hbar^{2}kq'}\!\left(\frac{\hbar^{2}k^{2}}{2m_a} - \hbar\omega\right)\,.
\end{align}
The lower limit of the first integration in Eq.~\eqref{piv<>_numerical} $q_{\min}$ is given by the condition $-1 \leq \xi_{0} \leq 1$, which originates from integration over the $\delta$ function in Eq.~\eqref{piv<>}
\begin{align}
  q_{\min}^{\mp}
  = &\,
  \frac{m_{a}}{\hbar^{2}k}\abs{\frac{\hbar^{2}k^{2}}{2m_{a}} \mp \hbar\omega}
  \,.
\end{align}

In order to numerically evaluate \eqa{piv<>_numerical}{piv<>_numerical_2}, we use nested adaptive Gauss-Legendre quadratures. For the principal value integration in $\zeta$, the integration points are chosen to be symmetrically distributed around the simple pole at $\zeta_{0}$. The second singularity at $a^2 = b^2$ is of an integrable form, but can still cause problems numerically. 


\subsection{First order self energy correction}
The second beyond RPA term in \eq{pi_series} accounts for the self energy of the particles arising from the influence of the self-consistent mean field. For this contribution consisting of two terms, one finds the following representation in the space-time domain
\begin{align}
  \label{piS0}
  \Pi_a^{\text{S}}(12)
  = &
  -i\hbar s_{a}
  {g_{a}^{0}}(21)\!
  \int\limits_{{\cal C}} \dif3\dif4\,
  {g_{a}^{0}}(13) \Sigma_{a}(34){g_{a}^{0}}(42) 
  \nonumber\\ &
  -i\hbar s_{a}
  {g_{a}^{0}}(12)\!
  \int\limits_{{\cal C}} \dif3\dif4\,
  {g_{a}^{0}}(23) \Sigma_{a}(34) {g_{a}^{0}}(41)
  \,.
\end{align}
Once again, the time integration is taken over the Keldysh contour. In \eq{piS0}, we have defined
\begin{align}
  \label{self_energy}
  \Sigma_{a}(34)
  = &~
  i\hbar\,{g_{a}^{0}}(34)V_{aa}^{\text{sc}}(34)
  \,.
\end{align}
Strictly, in \eq{self_energy}, we use only the first term of the GW or $V^{\text s}$ approximation. In order to be consistent in our perturbation expansion of the polarization function, the self energy \eqref{self_energy} is then given by the Hartree-Fock (HF) approximation \cite{kremp_book}
\begin{align}
  \label{self_energy_hf}
  \Sigma_{a}^{\text{HF}}(34)
  = &~
  i\hbar {g_{a}^{0}}(34)V_{aa}^{\text{sc}}(\rvec_{3} - \rvec_{4})\delta(t_{3} - t_{4})
  \,,
\end{align}
written here with respect to a statically screened interaction such as the Debye potential \eqref{debye_potential}. Note that \eq{self_energy_hf} relates to the exchange part only since the Hartree term may be neglected for charged particle systems due to charge neutrality.

Given the temporal structure of \eq{self_energy_hf}, there is no clear way to unambiguously treat the time ordering in Eq.~\eqref{piS0}. Thus, direct evaluation of the correlation functions using the Langreth-Wilkins rules is not possible. Instead, the self energy terms of the polarization function are computed starting from equations similar to the RPA term
\begin{align}
  \label{pishf}
  {\Pi_a^{\text{S}}}^{\gtrless}(12)
  =&~
  -i\hbar s_{a} {g_a^{\text{HF}}}^{\gtrless}(12){g_a^{0}}^{\lessgtr}(21)\nonumber\\
  &~
  -i\hbar s_{a} {g_a^{0}}^{\gtrless}(12){g_a^{\text{HF}}}^{\lessgtr}(21)
  \,,
\end{align}
where the HF correlation functions ${g_a^{\text{HF}}}^{\gtrless}$ contain only beyond RPA HF self energy contributions.
Self energy contributions are to be included in the spectral function
$a_a\ofkwt$ which directly determines the correlation functions. The spectral function can be obtained from \cite{kremp_book}
\begin{align}
a_a\ofkwt= &~ i\left[g_a^R\ofkwt-g_a^A\ofkwt\right]\,,
\end{align}
where the retarded and advanced Green's functions follow from the Dyson equation which is cut after the first iteration
\begin{align}
g_a^{\text{R/A}}(12)=&~ {g_a^0}^{\text{R/A}}(12)\\
+& \int\limits_{\cal C}\dif3\dif4\,
 {g_a^0}^{\text{R/A}}(13)\Sigma_a^{\text{HF}}(34){g_a^0}^{\text{R/A}}(42)\,.\nonumber
\end{align}
Transforming this equation into momentum-frequency space in local approximation leads to
\begin{align}
{g_a^{\text{R/A}}\ofkwt}=&~ \frac{1}{\hbar\omega-E_a^0(k)\pm i\epsilon}
+\frac{\Sigma_a^{\text{HF}}(\kvec;t)}{\big[\hbar\omega-E_a^0(k)\pm i\epsilon\big]^2}\,,
\end{align}
with $\epsilon\to 0^+$.
The spectral function for the pure self energy contribution then follows as
\begin{align}
a_a^{\text{HF}}\ofkwt= &~ \Sigma^{\text{HF}}(\kvec;t)
\frac{4\epsilon(\hbar\omega-E_0(k))}{\left[\big(\hbar\omega-E_0(k)\big)^2+\epsilon^2\right]^2}
\,.
\label{a_hf_only}
\end{align}
This expression is the first term in a perturbation series of the spectral function \eqref{gs_1} and can be interpreted as the self energy multiplied by the negative derivative of the $\delta$-function in Eq.~\eqref{gs_1}. The Fourier transform of \eq{self_energy_hf} can be shown to be
\begin{align}
  \label{self_energy_hf_fourier}
  \Sigma_{a}^{\text{HF}}(\kvec;t)
  = &~
  -\int\!\frac{\dif\qvec}{(2\pi)^{3}}\,
  f_{a}(\qvec;t)V_{aa}^{\text{sc}}(\kvec - \qvec;t)\,
  \nonumber \\
  = &
  \frac{Z_a^2e^2 k_C}{2\pi k}
  \int\limits_0^{\infty}\dif q\, q\, f_a(q;t)
  \nonumber\\
  & \times
  \ln\abs{\frac{(q + k)^{2}+\kappa_{e}^{2}(t)}{(q - k)^{2}+\kappa_{e}^{2}(t)}}\,.
\end{align}
Equation \eqref{self_energy_hf_fourier} represents a further generalization of the finite-wavelength screening wave number discussed in Ref.~\cite{CVFD_2015}. The more familar form of the HF self energy, in which the unscreened Coulomb potential appears, is recovered from \eq{self_energy_hf_fourier} by setting $\kappa_{e} = 0$.

The further evaluation proceeds similarly to the RPA case starting from Eq.~\eqref{pishf} with the HF particle propagator featuring the spectral function \eqref{a_hf_only}. The correlation functions for the self energy term of the polarization function is then given by
\begin{widetext}
\begin{align}
  \label{pis<>}
  \lefteqn{{\Pi_{a}^{\text{S}}}^{\gtrless}\ofkwt
  = 
  -4\pi^2 i s_{a}\!
  \int\!\!\frac{\dif\qvec}{(2\pi)^3}\int\!\!\frac{\dif\omega'}{2\pi}
  \Sigma_a^{HF}(\qvec;t)
  \frac{4\epsilon\big(\hbar\omega'-E_a^0(\qvec)\big)}
  {\left[\big(\hbar\omega'-E_a^0(\qvec)\big)^2+\epsilon^2\right]^2}
  }&\\
  &\times
  \left\{f_a^{\gtrless}(\hbar\omega'+\hbar\omega;t)f_a^{\lessgtr}(\omega';t) 
  \delta\left(\hbar\omega'+\hbar\omega-E_a^0(\qvec+\kvec)\right)\right.&\nonumber\\
 &\left. \qquad+
  f_a^{\gtrless}(\omega';t) f_a^{\lessgtr}(\hbar\omega'-\hbar\omega;t)  
   \delta\left(\hbar\omega'-\hbar\omega-E_a^0(\qvec-\kvec)\right)
  \right\}\nonumber\,.
\end{align}
\end{widetext}
In \eq{pis<>}, one of the four integrations is trivial and a further one can be performed immediately with the help of the $\delta$-functions. The final result as it will be used for numerical evaluations is
\begin{widetext}
\begin{align}
  \label{pis<>_num}
  {\Pi_{a}^{\text{S}}}^{\gtrless}(k,\omega;t)
  = &
  -\frac{i s_{a}}{(2\pi)^2}\!
  \int\limits_0^{\infty}\!\!q^2\dif q\int\limits_{-1}^{1}\!\!\dif x\,
  \Sigma_a^{HF}(q;t)
  \left\{f_a^{\gtrless}(\qvec+\kvec;t)
  f_a^{\lessgtr}(E_a^0(\qvec+\kvec)-\hbar\omega;t) 
  \frac{4\epsilon\left(E_a^0(k)-\hbar\omega-\frac{2qkx}{2m_a}\right)}
  {\left[\big(E_a^0(k)-\hbar\omega-\frac{2qkx}{2m_a}\big)^2+\epsilon^2\right]^2}
  \right.\nonumber\\
  &+\left.
  f_a^{\gtrless}(E_a^0(\qvec-\kvec)+\hbar\omega;t) f_a^{\lessgtr}(\qvec-\kvec;t)  
    \frac{4\epsilon\left(E_a^0(k)+\hbar\omega+\frac{2qkx}{2m_a}\right)}
  {\left[\big(E_a^0(k)+\hbar\omega+\frac{2qkx}{2m_a}\big)^2+\epsilon^2\right]^2}
  \right\}\,.
\end{align}
\end{widetext}
In order to evaluate \eq{pis<>_num}, one needs to choose the free parameter $\epsilon$ sufficiently small and run convergence tests. This has been successfully done without any problems in the current work.

The derivation of the retarded polarization function from Eq.~\eqref{pis<>} proceeds using the Kramers-Kronig relation \eqref{pir}. Specifically, one obtains two distinct contributions to the retarded self energy term despite their Feynman graphs being topologically equivalent (see \fig{pi_expansion}):
\begin{widetext}
\begin{align}
  \label{pisr}
  {\Pi_a^{\text{S}}}^{\ret}\ofkwt
  = &~
  s_{a}\int\!\frac{\dif\qvec}{(2\pi)^3}\,
  \Sigma_{a}^{\text{HF}}(\qvec;t)
  \left\{
  \frac{f_a(\qvec+\kvec;t) - f_a(\qvec;t)}
  {\sqr{\hbar\omega+i\epsilon - E_{a}^{0}(\qvec+\kvec) + E_{a}^{0}(\qvec)}^2}
  +\frac{\partial f_a(\qvec;t)/\partial E_{a}^{0}\ofq}
  {\hbar\omega+i\epsilon - E_{a}^{0}(\qvec+\kvec) + E_{a}^{0}(\qvec)}
  \right\}
  \nonumber\\
  +&~
  s_{a}\int\!\frac{\dif\qvec}{(2\pi)^3}\,
  \Sigma_{a}^{\text{HF}}(\qvec;t)
  \left\{
  \frac{f_a(\qvec - \kvec;t) - f_a(\qvec;t)}
  {\sqr{\hbar\omega+i\epsilon - E_{a}^{0}(\qvec) + E_{a}^{0}(\qvec - \kvec)}^2}
  -\frac{\partial f_a(\qvec;t)/\partial E_{a}^{0}\ofq}
  {\hbar\omega+i\epsilon - E_{a}^{0}(\qvec) + E_{a}^{0}(\qvec - \kvec)}
  \right\}
  \,.
\end{align}
\end{widetext}
Comparison with the result obtained using the imaginary time Matsubara method is possible recognising that 
$\partial f_{a}(\qvec)/\partial E_{a}^{0}(\qvec) = -\beta f_{a}(\qvec)[1 - f_{a}(\qvec)]$ for an equilibrium Fermi-Dirac distribution. It is immediately clear that the first term of Eq.~\eqref{pisr} is exactly equivalent to the result given by DeWitt {\em et al.} \cite{DSSK_1995}. The second term of Eq.~\eqref{pisr} is not given in the latter paper, instead the first term is multiplied by two to account for the second diagrammatic contribution. DeWitt {\em et al.} were aiming to derive corrections to the EOS in the non-degenerate limit \cite{DSSK_1995}.
Indeed, analysis of \eq{pisr} for non-degenerate equilibrium systems for $k\to0$ and $\omega = 0$, recovers their important well-known result (see Appendix B)
\cite{DSSK_1995}
\begin{align}
    \label{dewitt_result}
    {\Pi_{a}^{\text{S}}}^{\ret}(\kvec \to 0,0)
    \stackrel{D_{a} \ll 1}{=} &\,
    -\frac{n_{a}\beta\rnd{\kappa_{\text{D}a}\lambda_{a}}^{2}}{2s_{a}}
    \,,
\end{align}
where $\kappa_{\text{D}a}^{2} = Z_{a}^{2}e^{2}n_{a}\beta4\pi k_{\text{C}}$ and $\lambda_{a}^{2} = \beta\hbar^{2}/m_{a}$ are the square Debye wave number and de Broglie wavelength, respectively. The retarded function related to the vertex correction gives an identical result under these special limits and conditions.

For the general case however, the correlation functions are required to fulfill the general time symmetry condition
\begin{align}
  \label{time_reversal_symmetry}
  \Pi^{>}(\kvec,-\omega;t) 
  = 
  \Pi^{<}\ofkwt
  \,.
\end{align}
Again, this holds for all combinations of species labels. It is trivial to demonstrate that \eq{pis<>_num} obeys the condition \eqref{time_reversal_symmetry}. In contrast, the time symmetry condition is not fulfilled if the retarded self energy correction is given by constructing the correlation functions from taking twice either of the two terms in \eq{pisr}. 

Another important comparison for our result is provided by the results of Holas et al. \cite{HAS_1979}. In the latter, the ground state ($T = 0$) electron gas was investigated for arbitrary frequencies. Holas {\em et al.} obtained two distinct contributions to the self energy, similar to our \eq{pisr}. However, the terms featuring derivatives of Wigner distribution functions are missing in the expressions given by Holas {\em et al.}, but these give finite contributions even for $T=0$ since the derivative of the step function yields a $\delta$-function. 

In evaluating the retarded self energy contribution to the polarization function \eq{pisr}, we numerically solve the Kramers-Kronig relation (\ref{pir})  based on the result for the self energy correlation functions \eq{pis<>_num}, or choose a value for the parameter $\epsilon$ and evaluate \eq{pisr} directly. Naturally, properly converged (with respect to $\epsilon$) calculations show identical results using either method.

\begin{figure}[t]
  \includegraphics[width=0.45\textwidth,clip=true]{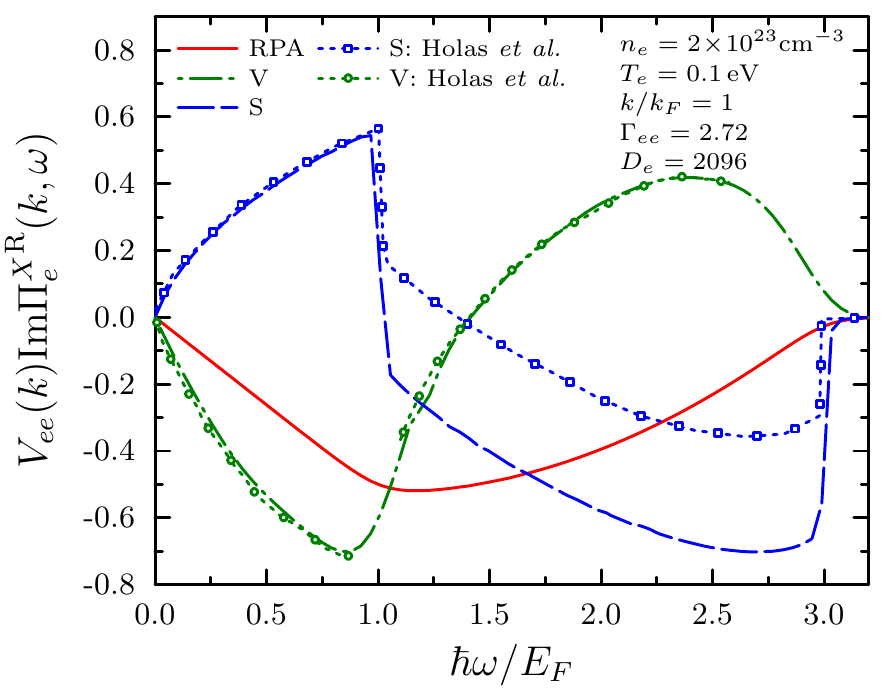}
  \caption{
    Results for the RPA (solid red curve), (linearized) self energy (dashed blue curve) and vertex (dash-dotted green curve) contributions to the imaginary part of the retarded polarization function in a highly degenerate equilibrium electron gas $n_e=2\!\times\!10^{23}\,\unit{cm^{-3}}$, $T = 0.1 \,\unit{eV}$, $k = k_{\text{F}} = 0.964\,a_{\text{B}}^{-1}$. Results from Holas et al. \cite{HAS_1979} are shown for comparison (dotted curves with markers).
    }
  \label{sv_t=0}
\end{figure}

\begin{figure}[t]
  \includegraphics[width=0.45\textwidth,clip=true]{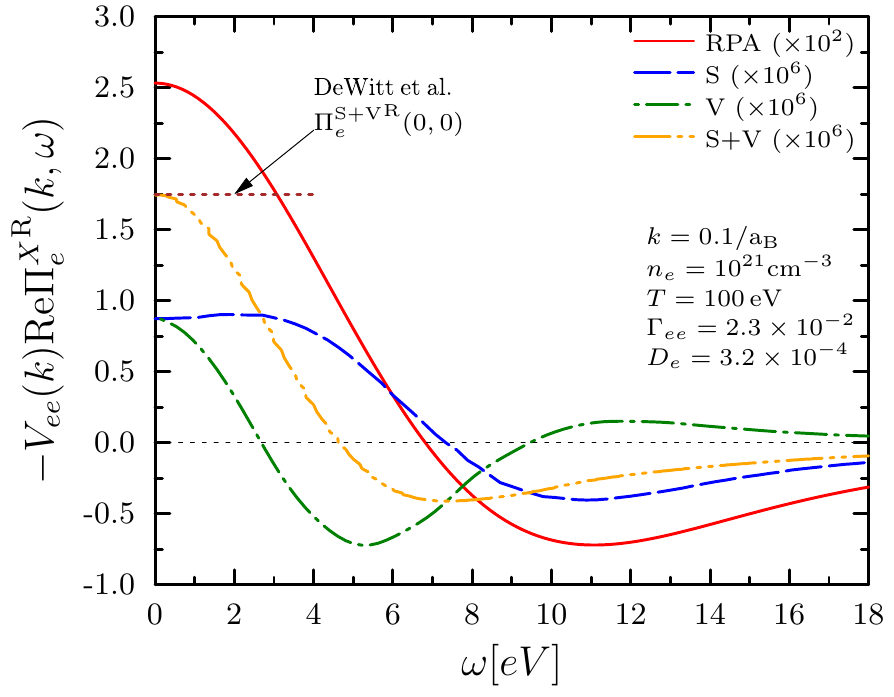}
  \caption{
    Results for the real part of the retarded vertex function (dash-dotted green line) and self energy (dashed blue line) in a non-degenerate electron gas at $n_e=10^{21}\,\unit{cm^{-3}}$, $T=100\,\unit{eV}$, $k=0.1 \, a_{\text{B}}^{-1}$ in equilibrium. Current results (dash-dot-dotted line is the sum of vertex and self energy term) are compared to results from DeWitt et al. via Eq.~(\ref{dewitt_result}) (dashed brown marker) \cite{DSSK_1995}.
    }
    \label{sv_thigh}
\end{figure}
Figures \ref{sv_t=0} and \ref{sv_thigh} provide a comparison between the expressions obtained in this work and literature values for the case of thermal equilibrium.
The dynamic vertex and self-energy terms for a strongly coupled degenerate electron gas are shown in Fig.~\ref{sv_t=0} together with results by Holas et al. \cite{HAS_1979}. To avoid numerical issues arising from evaluating the model at zero temperature, we use a temperature of $T = 0.1\,\unit{eV}$. The degeneracy and coupling parameters for this example are $D_{e} = n_{e}\Lambda_{e}^{3} = 2096$ and $\Gamma_{ee}=\langle E_{pot}\rangle/\langle E_{kin}\rangle = 2.72$. We find excellent agreement with the results of Holas {\em et al.} for the vertex term \citep{HAS_1979}, but a significant difference for the self energy correction is seen. We attribute this discrepancy to the lack of the second terms in the retarded function in Eq.~(2.7) Holas {\em et al.}.

For the case of a non-degenerate electron gas as shown in Fig.~\ref{sv_thigh}, the long-wavelength limit ($k\to 0$) of the static ($\omega = 0$) retarded polarization function has a known analytic result \citep{DSSK_1995}. This limit Eq.~(\ref{dewitt_result}) is shown to be well-reproduced by our results in Fig.~\ref{sv_thigh}. We further confirm that the real parts of the retarded self energy and vertex terms give the same value in the limits of small frequencies and momenta, as expected, despite showing different dynamic behaviours.

\section{Results for the non-equilibrium structure\label{sec6}}
\subsection{Comparison with equilibrium results}

In order to compare the present work to results from the literature, we are necessarily restricted to thermal equilibrium. In this case, higher order contributions beyond RPA can be included into the structure factor and dielectric function in a variety of different ways \cite{RRRW_2000,AADD_2010,VGK_2013,MDG_2012}. In this work, we consider the fully dynamic properties of the correlations, but we are limited to systems with weak interactions. A well-known alternative approach makes use of dynamic local field corrections (LFCs). Static LFCs have many known properties in the classical non-degenerate and $T=0$ (fully degenerate) limits and for the ($\omega=0$) and high-frequency ($\omega\to\infty$)cases \cite{mahan_book,kremp_book,STLS_1968,IKP_1984}. For intermediate temperatures or arbitrary frequencies, interpolation formulas or expansions are often used \cite{D_1986,HR_1987,HK_1991,SB_1993,GRHG_2007,FWR_2010,VDTG_2012}. 

For a direct comparison of the DSF, we consider conditions of interest to experiments in the WDM regime. Specifically, we choose the example of solid density aluminium, i.e.~$n_{e}=1.8\!\times\!10^{23}\,\unit{cm^{-3}}$, heated to $10\,\unit{eV}$, e.g.~using a short-pulse laser-produced proton beam \cite{HGJG_2012,PGGV_2010}. Under these conditions, both the electron-electron coupling and degeneracy are moderate; $\Gamma_{ee} = 1.22$ and $D_{e} = 1.9$. Subsequently, no single theory of the polarization function is valid without some restriction. We consider wave numbers between $k = 0.5-1.5\,a_{\text{B}}^{-1}$ in order to demonstrate the effect of correlations over a range collectivity parameters $\alpha = \kappa_{e}/k$. As expected, all curves join the RPA result for high energies but different predictions for the electron structure are given for small energies. The current approach of using dynamic vertex- and self energy contributions featuring a screened potential agrees well with results from static LFCs ($\omega=0$). The vertex and self energy expansion featuring a Coulomb potential gives the highest values. 
This is an interesting comparison and the good proximity of our result (using a Debye potential) and of the LFCs shows that the fits of temperature dependence and frequency dependence used to calculate the LFCs work reasonably well in this regime.

\begin{figure}[t]
  \includegraphics[width=0.45\textwidth,clip=true]{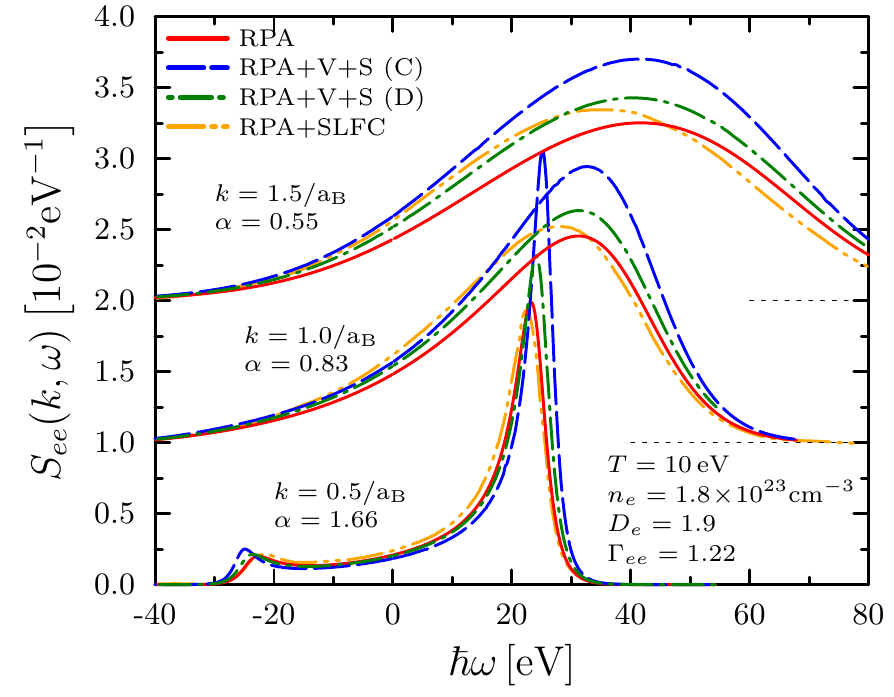}
  \caption{
   Comparison of the DSF of an electron gas at the conditions expected in warm dense aluminium using the first-order corrections to the polarization function: $n_{e}=1.8\times 10^{23}\,\unit{cm^{-3}}$, $T=10\,\unit{eV}$. Several wave numbers $k = 0.5-1.5\,a_{\text{B}}^{-1}$ are considered as well as the effect of using Coulomb and Debye potentials.
    }
    \label{see_gk}
\end{figure}
Including the vertex and self energy terms within the framework presented in this work using an unscreened Coulomb potential gives a considerably different electronic structure than using a screened potential. This extends to the location and width of the plasmon peak. The different approximation of electron-electron interactions is responsible and it is clear that compensation effects occur and that screening reduces the correlation strength and therefore the deviation from RPA. 

For realistic situations, electron-ion collisions are still to be included. This can be done in an efficient way in equilibrium using the extended Mermin approach of Fortmann et al. \cite{FWR_2010}. In non-equilibrium, the evaluation of all the second order terms, including the cross species contributions responsible for electron-ion collisions, is needed and therefore a priority for future investigations.

\subsection{Non-equilibrium example: FEL-pumped electron distribution}

In order to provide an informative example of the effects of non-equilibrium in a realistic system of interest, we consider probing an iron WDM state as discussed in Ref. \cite{WNKD_2013} with x-rays from a high-intensity free-electron laser, such as LCLS or the Euro-XFEL machine. It is well known that under such a high flux of energetic photons, excitations due to photo-ionization, Auger decay and collisional ionization leads to strongly non-equilibrium distributions functions featuring a quasi-thermal high energy tail and several roughly Gaussian-shaped bumps \cite{FBDD_2010,MZFG_2011}. Whilst the model outlined in this work is clearly suitable for distribution functions with arbitrarily complicated shapes, it is often reasonable to consider simple analytic models which adequately capture the relevant features. One such example is the \lq{bump-on-hot-tail}\rq~model distribution function \cite{CG_2011,CVWG_2012}
\begin{align}
  \label{nedf}
  f_{e}(k)
  =&\,
  A_{c} \sqr{\exp\rnd{\beta_{c}\rnd{\frac{\hbar^{2} k^{2}}{2m_{e}}-\mu_e}} + 1}^{-1}
  \nonumber\\ & 
  +
  A_h \exp\rnd{-\beta_h\frac{\hbar^{2}k^{2}}{2m_{e}}}
  \nonumber\\ & 
  +
  A_b \exp\rnd{-\beta_b\frac{(\hbar k-p_b)^2}{2m_{e}}}\,.
\end{align}

In order to accentuate non-equilibrium effects, we consider a relatively large fraction of electrons moved from the Fermi-shaped bulk component into high-energy, non-thermal states. Specifically, the total number density of free electrons is fixed at $n_{e} = 10^{24}\,\unit{cm^{-3}}$ and the total fraction in the bulk $A_{c} = n_{c}/n_{e}$ is set at $A_{c} = 0.5$, with the remainder being located in the bump. The momentum offset of the bump is set at $p_{b} = 5 \hbar a_{\text{B}}^{-1}$, which gives an energy in the range of K-shell Auger emission. The amplitudes of the various components are adjusted to give the correct total density upon integrating over the momentum \cite{kremp_book}.

\begin{figure}[t]
  \includegraphics[width=0.45\textwidth,clip=true]{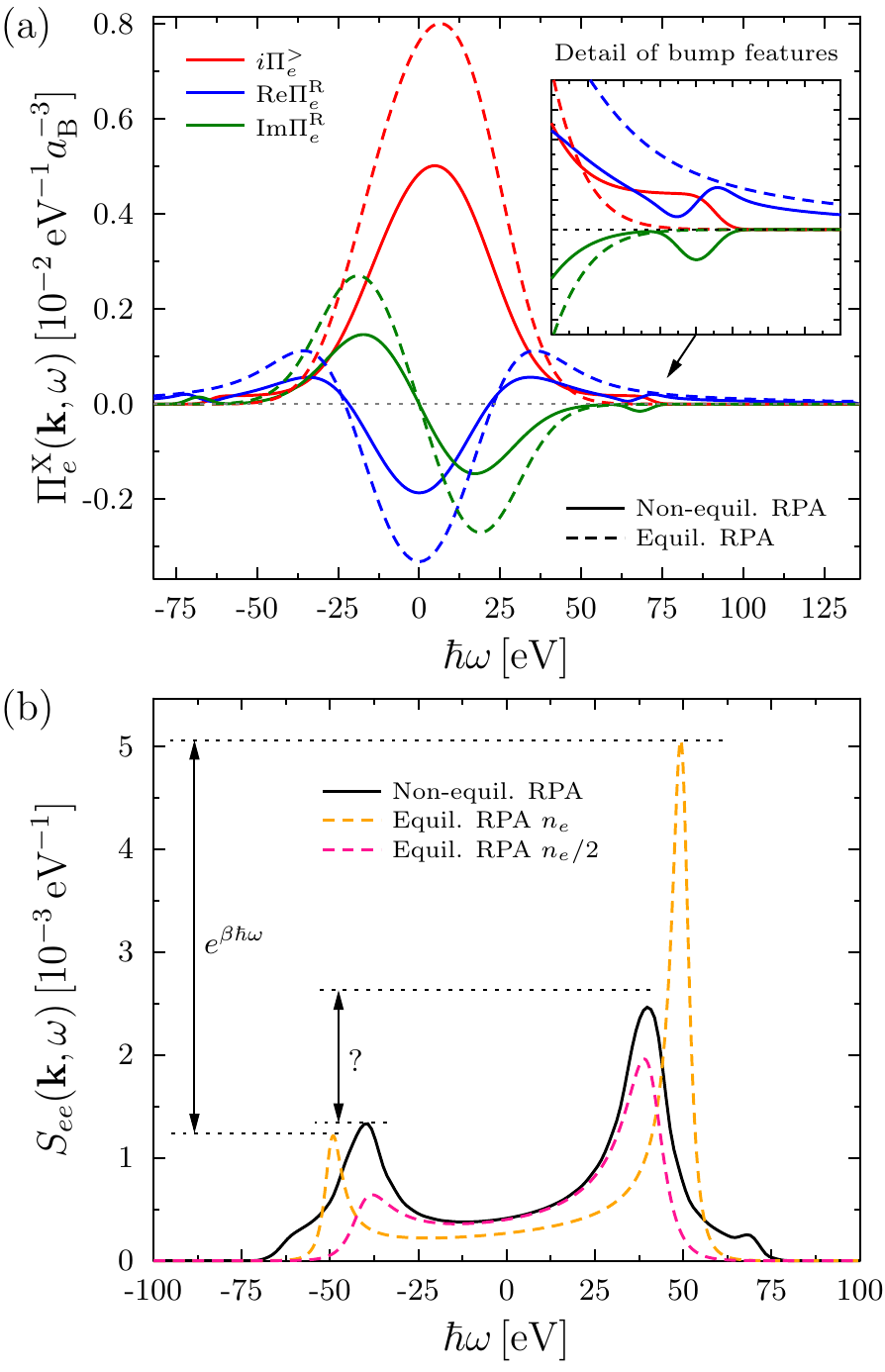}
  \caption{
    Structure in a non-equilibrium electron gas in RPA with $n_e = 10^{24}$cm$^{-3}$ where half the density is in a distribution with $T_{c} = 4\times 10^5\,\unit{K}$ and one half of the density is in a Gaussian bump with $T_b = 10^4\,\unit{K}$ and $p_b = 5\, \hbar /a_{\text{B}}$. The wave number of the density fluctuations is $k = 0.5\,a_{\text{B}}^{-1}$. (a): Correlation and related retarded functions for the polarization function. (The equilibrium result is for $n_e$.) (b): Dynamic structure factor. 
    }
  \label{see_detail_bal}
\end{figure}

\begin{figure}[t]
  \includegraphics[width=0.45\textwidth,clip=true]{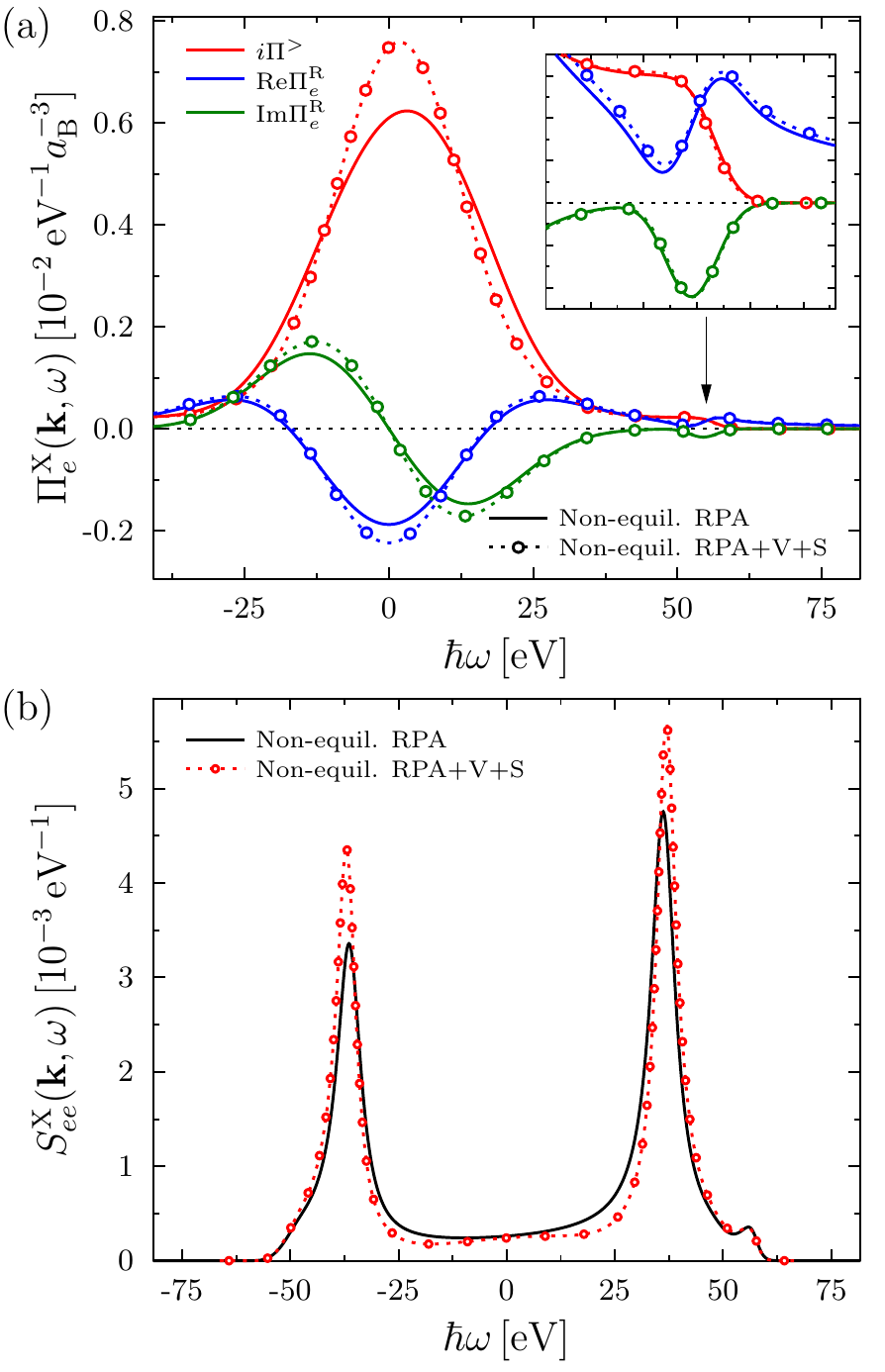}
  \caption{
    The change in the non-equilibrium structure due to the inclusion of vertex and self energy terms for the same conditions as in Fig. \ref{see_detail_bal} but a wave number of $k = 0.4\,a_{\text{B}}^{-1}$. (a): Correlation and retarded functions for the polarization function $i\Pi_{e}^{>}(\kvec,\omega)$ in RPA and including vertex and self energy. (b): The dynamic structure factor in a non-equilibrium electron gas in RPA and including vertex and self energy corrections.
    }
  \label{see_detail_bal_2}
\end{figure}

The effect of the non-equilibrium distribution \eqref{nedf} on the different polarization functions in RPA are displayed in the top panel of Fig.~\ref{see_detail_bal}. The general trend is towards a reduction in the magnitude of the quantities as a function of frequency due to the lower density of electrons in the bulk of the distribution compared to equilibrium. Moreover, small features around $\hbar\omega=\pm 70 \, \unit{eV}$ appear due to the bump. These are shown in greater detail for positive frequencies in the inset panel.

The bottom panel of Fig.~\ref{see_detail_bal} presents the results for the DSF resulting from this non-equilibrium state, which shows several distinct differences in comparison to the equilibrium calculation. Firstly, we note that the plasmon peaks are strongly down-shifted. This is due to the plasmon dispersion relation being largely determined by the density of the bulk component for the value of the momentum transfer studied in this example. Indeed, an equilibrium calculation at the same temperature but half the electron density gives peaks at similar frequency shifts. The location, but not the height or width, can be well approximated by this equilibrium calculation.

We also see a significant difference in the relative amplitudes of the plasmon peaks. In equilibrium, this ratio is given by the detailed balance relation, which is intrinsically related to the temperature according to 
\begin{align}
  \label{detailed_balance}
  S_{ee}(-\kvec,-\omega) 
  = &\
  e^{-\beta\hbar\omega}
  S_{ee}(\kvec,\omega). 
\end{align}
Equation \eqref{detailed_balance} reflects the relative lack of availability of free states for electrons to occupy after scattering a photon, which increases as the system tends toward higher degeneracy. The same simple relationship does not exist for non-equilibrium states in general. Discrepancies between the plasmon damping and detailed balance may therefore act as a sensitive experimental indicator of departures from equilibrium.

Interestingly, the high-frequency features of the polarization functions becomes substantially more prominent in the DSF, while at the same time step-like wings extend beyond the plasmon peaks. These features arise since the magnitude of the dielectric function is dominated by the cold bulk, which decays rapidly at high frequencies, whilst the correlation function shows significant structure due to the electrons in the non-thermal bump. The presence of these features is another characteristic signature which may be used to infer non-equilibrium physics in experimental data.


\begin{figure}[t]
  \includegraphics[width=0.45\textwidth,clip=true]{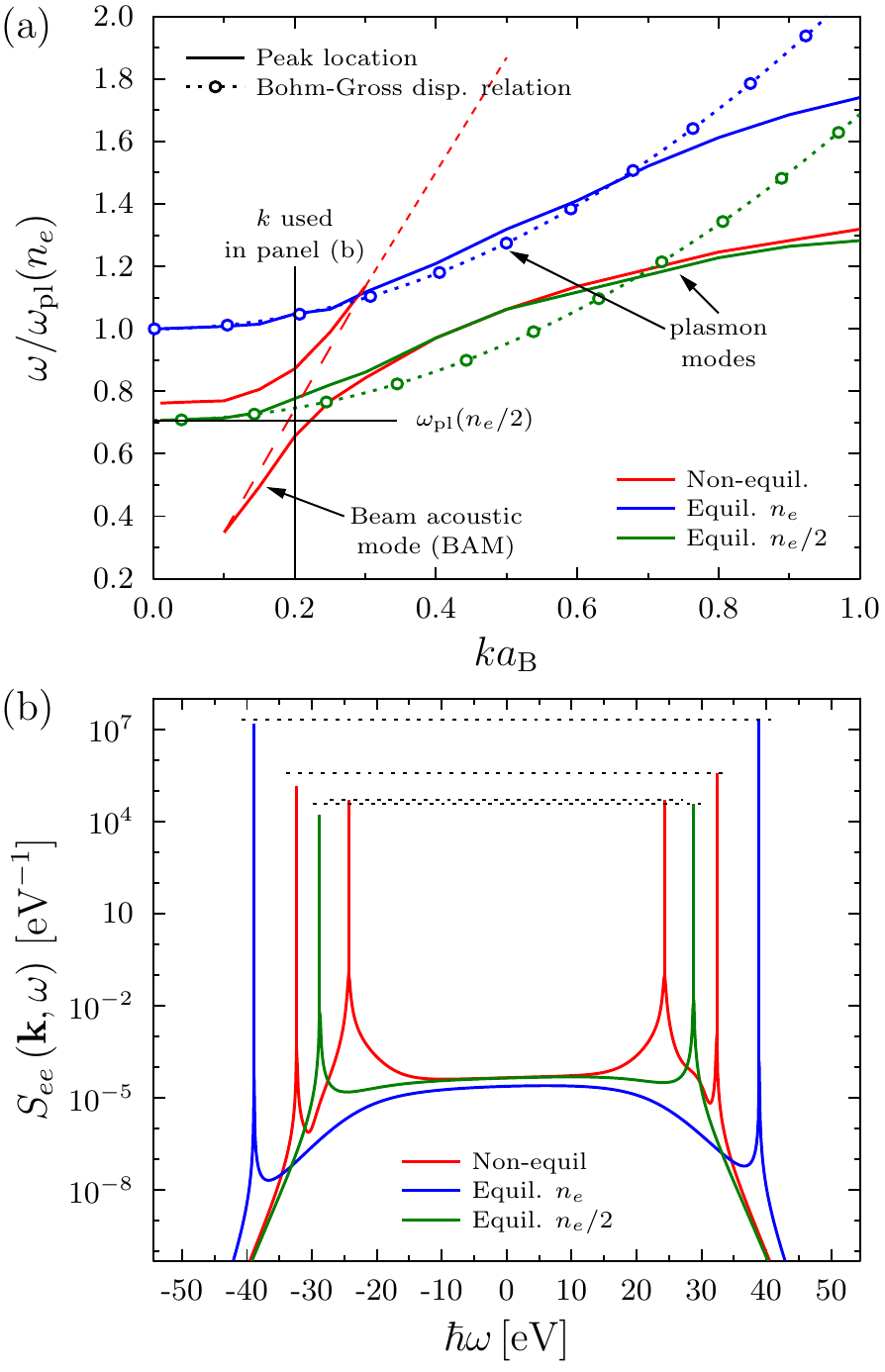}
  \caption{
    Dispersion relation of longitudinal plasma excitations for the parameters of Fig. \ref{see_detail_bal} (the curves are derived from the maximum of the spectral function, {\em} not from the zeros of the dielectric function). The discontinuity in the dispersion curve for the non-equilibrium case is due to an interaction of the plasmon and the beam acoustic excitation. The red long dashed line connecting the end points of the discontinuity in the BAM represents the dispersion of an unperturbed acoustic mode (sound speed $c_s=4.74\times 10^5 \, \unit{ms^{-1}}$) and is shown as a guide for the eye. The read short dashed line indicates for which $k$-values this second maximum in the spectral function exists.
    }
  \label{see_disp}
\end{figure}

Including the first order correlation corrections results in significant relative increases in the magnitudes of the different polarization functions compared to RPA (see top panel of Fig.~\ref{see_detail_bal_2}). However, such clear difference are not observed in the dynamic structure, which shows only increases in the amplitudes at the resonance frequencies of the plasmon peaks. This can be understood since the components of the dielectric function do not change significantly close to where $\text{Re}\,\varepsilon_{ee}(\kvec,\omega) = 0$. It is interesting to note that the observed amplification and stabilisation of the plasmons is the opposite behavior compared to equilibrium, where correlations and exchange usually broaden the resonances. In contrast, the step-like non-equilibrium shoulder is not strongly affected by the addition of the vertex and self energy terms.


The dispersion of the collective excitations of the non-equilibrium system considered in the present example is shown in panel (a) of Fig.~\ref{see_disp}. For small momenta, only a single excitation (the plasmon) exists, the frequency of which is approximately given by the effective plasma frequency resulting from the electron density contained in the cold bulk of the distribution function. The same is true for the highest momenta shown. The dispersion roughly follows a Bohm-Gross-like relation in the long-wavelength limit and deviates from it once the damping enters the non-linear regime \cite{kremp_book}. For wave numbers between $0.1 \leq ka_{\text{B}} \leq 0.3$, a second collective excitation, known as the beam acoustic mode (BAM), is present. The BAM has been described previously using different methods and for a range of different conditions \cite{OM_1968,KMBC_2005,VYDB_2005,CG_2011}. Due to the screening interaction between the plasmon mode and the BAM, which is mediated by the dielectric function, the dispersion branches do not cross; the plasmon mode gets pushed to higher frequencies and the BAM gets pushed to lower frequencies. The result is that the upper branch ends up being the BAM excitation even though it was of plasmon character for small wave numbers. Similarly, the lower mode starts out as a BAM but changes its character before the upper branch ceases to exist and re-emerges as the plasmon mode. The dispersion relation for small wave numbers in the BAM case can be described by a modified Bohm-Gross relation taking into account the high energy tail of the distribution function which leads to the larger prefactor in the $k^2$-behavior \cite{CG_2011}.

If the wave number used to study the DSF in Figs.~\ref{see_detail_bal} and \ref{see_detail_bal_2} is reduced to a value where both the regular plasmon mode and the BAM exist (see vertical marker in panel (a) of Fig.~\ref{see_disp}), strong spectral signatures in the excitation spectrum emerge. As shown in panel (b) of Fig.~\ref{see_disp}, the resulting structure is quintessentially different to the previous example, where the plasmon peaks of the non-equilibrium system are approximated well by an equilibrium system with half the density. This is due to the occurrence and interaction of plasmon and BAM. Furthermore, whereas the amplitudes of the plasmon-like peaks are asymmetric (remnants of detailed balance), the BAM excitations are symmetrical. In this case, the changes to the dynamic structure due to the inclusion of vertex and self energy contributions are not so easily described. Particularly in the frequency range of the BAM, vertex and self energy terms have opposite signs. Thus, one may conclude from this example that in a general non-equilibrium situation the influence of correlations and exchange on damping and location of collective modes cannot be predicted ad hoc, not even qualitatively. Instead, the only reliable method for understanding the DSF is to calculate it directly from the distribution function.


\section{Decomposition of the dynamic structure factor in non-equilibrium\label{sec3}}
The decomposition of the equilibrium electron DSF according to Chihara \cite{C_1987,C_2000} is the basis for the analysis of experimental x-ray scattering spectra and a cornerstone of interaction between theory and experiment (see, e.g., \cite{WVGG_2011,VG_2015} amongst other works). The Chihara formula uses the chemical picture to distinguish between free and bound electron contributions in a semi-classical framework. The resulting three terms are:~1) The \mbox{(free-)} electron gas term describing the high-frequency response of the electrons in the continuum,~2) The ion term describing low-frequency (quasi-elastic) scattering from bound states and the screening cloud of free electrons which surrounds the ions,~3) The bound-free term describing Raman-like transitions from bound states into the continuum. 

The utility of the Chihara formula is based on its identification of the different correlation contributions to the total structure factor, whilst not attempting to describe in detail each contribution from first principles. The different contributions of the DSF can then be evaluated using a variety of different models or techniques and have recently focused heavily on incorporating ab initio simulation results (see, e.g., \cite{VG_2015,FLDG_2015}). In comparison, the fully quantum mechanical approach in the physical picture requires direct evaluation of the polarization functions and, thus, is currently significantly restricted in practice \cite{BSDH_2016}, or relies on the Born-Oppenheimer approximation \cite{WSGR_2017}. Treating the excitations of the electron gas on the same level as Raman transitions and rigorously describing the low-frequency response of the ions under strongly coupled conditions are particularly challenging within the present framework.

The original Chihara formula is also based on the assumption of equilibrium. Naturally, a decomposition in a similar spirit to the Chihara formula would be of considerable use for the general case of non-equilibrium states. 
Such a formalism would allow simplifications to be rigorously derived for special cases of interest, such as two-temperature states or systems with one or more non-equilibrium component. The framework presented in this paper is ideally suited for this application. 

We start by defining a generalized screening cloud (or generalized form factor) in the space and time domain $\rho(\rvec_{1}t_{1},\rvec_{2}t_{2})$ via the electron-ion and ion-ion correlation functions
\begin{align}
  \label{leitot}
  L_{ei}(\rvec_{1}t_{1},\rvec_{2}t_{2})
  = &\,
  \int\dif\rvec_{3}\int\limits_{\cal C}\! \dif t_3 \,\,\rho(\rvec_{1}t_{1},\rvec_{3}t_{3}) 
  L_{ii}(\rvec_{3}t_{3},\rvec_{2}t_{2})
  \,.
\end{align}
As throughout the whole paper, we operate in the physical picture and do not explicitly distinguish between bound and free electrons. Consequently, the electron-ion density fluctuation correlation function in \eq{leitot} includes all electrons. Thus, the generalized screening cloud $\rho$ includes bound electrons, free electrons, as well as dynamic effects like bound-free transitions. 

After transferring from the Keldysh contour to the physical time axis, enforcing the local approximation, and Fourier transforming with respect to the microscopic difference variables, we obtain for the correlation and retarded/advanced functions for $L_{ei}$
\begin{align}
  \label{L_ei_>_def}
  L_{ei}^{>}
  = &\,
  \rho^{>} L_{ii}^{\adv} + \rho^{\ret} L_{ii}^{>}
  \,,
  \\
  \label{L_ei_R/A_def}
  L_{ei}^{\retadv}
  = &\,
  \rho^{\retadv} L_{ii}^{\retadv}
  \,.
\end{align}
As usual, all the terms depend on the variables $\cly{\kvec,\omega,t}$.
Using \eqa{L_ei_>_def}{L_ei_R/A_def} together with the appropriate relations (\ref{app_L_ab_R/A}), we subsequently find
\begin{widetext}
\begin{align}
  \label{rg}
  \rho^{>}
  = &\,
  \frac{1}
  {|(1 - \tcal_{ee}^{\ret})\qcal_{ii}^{\ret}|^2}
  \Big\{
  \qcal_{ei}^{>} (1-\tcal_{ii}^{\adv}) \qcal_{ii}^{\ret}
  (1-\tcal_{ee}^{\adv})
  +\qcal_{ei}^{\adv} (1-\tcal_{ii}^{\adv}) \qcal_{ii}^{\ret}\tcal_{ee}^{>}
  \nonumber\\
  &
  -\qcal_{ei}^{\ret} (1-\tcal_{ii}^{\adv}) \qcal_{ii}^{>} (1-\tcal_{ee}^{\adv})
  \vphantom{\frac{1}{1}}
  -\qcal_{ei}^{\ret}\tcal_{ii}^{>}\qcal_{ii}^{\adv} (1-\tcal_{ee}^{\adv})
  \Big\}
  \,,
\end{align}
\end{widetext}
for the greater correlation function and
\begin{align}
  \label{rra}
  \rho^{\retadv}
  = &\,
  \frac{(1 - \tcal_{ii}^{\retadv})\qcal_{ei}^{\retadv}}
  {(1 - \tcal_{ee}^{\retadv})\qcal_{ii}^{\retadv}}
  \,,
\end{align}
for the retarded/advanced functions.  Applying the DPA, the familiar expressions for the screening cloud immediately follow
\begin{align}
  \rho^{\text{X}}
  =
  \lcal_{ee}^{\text{X}}V_{ei}\
  \,,
\end{align}
where $\text{X}$ again stands for $>$ or $\retadv$, as required. Under this approximation, the functions $\rho^{>}$ and $\rho^{\retadv}$ describe the usual screening cloud of free electrons in non-equilibrium.
 
If the off-diagonal elements of the  polarization functions are retained, as in \eqa{rg}{rra}, the generalized screening cloud includes the bound-electron (ionic) form factor, the screening cloud of free electrons, and also in principle any bound-free transitions. The actual evaluation of the ion form factor or bound-free transitions in non-equilibrium depends on the ability to either solve the screened ladder for the polarization function or to incorporate solutions of the time-dependent Schr\"{o}dinger equation into the present formalism.

In the physical picture, the Chihara-like decomposition of the DSF may be obtained using the ansatz
\begin{align}
  \label{neq_chihara_ansatz}
  L_{ee}(\rvec_{1}t_{1},\rvec_{2}t_{2})
  = &\,
  L_{ee}^{\dagger}(\rvec_{1}t_{1},\rvec_{2}t_{2})
  \nonumber\\ 
  & + \int\dif\rvec_{3}\int\limits_{\cal C}\! \dif t_{3} \,
  L_{ei}(\rvec_{1}t_{1},\rvec_{3}t_{3})\rho(\rvec_{3}t_3,\rvec_{2}t_2)
  \,,
\end{align}
with the electron-ion term given by Eq.~\eqref{leitot}. The purpose of introducing the two distinct contributions in \eq{neq_chihara_ansatz} is to separate the contributions from electrons that respond on high-frequencies, through the first term $L_{ee}^{\dagger}$, to those which are connected with the dynamic ion structure in the second term. In doing so, we do not weaken the rigor of the physical picture, nor do we specifically invoke the Born-Oppenheimer approximation. Instead, the frequency separation of the various contributions occurs naturally due to the strongly decaying nature of the ion structure factor at high-frequencies. 

The corresponding correlation and retarded/advanced functions in Fourier space follow from \eq{neq_chihara_ansatz} as
\begin{align}
  L_{ee}^{>}
  = &\,
  {L_{ee}^{\dagger}}^{\!\!>}+\rho^{>} L_{ii}^{\adv} \rho^{\adv}
  +\rho^{\ret} L_{ii}^{>} \rho^{\adv}+\rho^{\ret} L_{ii}^{\ret} \rho^{>}\,,
  \nonumber\\
  = &\,
  {L_{ee}^{\dagger}}^{\!\!>}+\rho^{>}\left(L_{ei}^{\adv}+L_{ei}^{\ret}\right)
  +\left|\rho^{\ret}\right|^2 L_{ii}^{>}\,,
  \nonumber\\
  = &\,
  {L_{ee}^{\dagger}}^{\!\!>} 
  + 2\rho^{>} \text{Re}\,L_{ei}^{\ret}
  +\left|\rho^{\ret}\right|^2 L_{ii}^{>}\,.  
\end{align}
Here, relation (\ref{L_ei_R/A_def}) was used to obtain the second line. We can use Eq.~\eqref{see} to immediately derive the corresponding dynamic structure factors
\begin{align}
  \label{noneq_chihara}
  S_{ee}\ofkwt 
  = &\,
  S_{ee}^{\dagger}\ofkwt
  \nonumber\\
  &+\frac{i\hbar}{\pi}\rho^{>}\ofkwt\mbox{Re}L_{ei}^{\ret}\ofkwt
  \nonumber\\
  &+\left|\rho^{\ret}\ofkwt\right|^2 S_{ii}\ofkwt\,,
\end{align}
which is the desired non-equilibrium generalization of the Chihara formula in the physical picture. This non-equilibrium generalisation of the Chihara formula offers the same degree of utilitarianism as the original equilibrium framework since one may supplement each component of the total DSF using complementary theoretical techniques. On the other hand, we present here the exact expressions for all the constituent terms on the basis of the polarization function.
 
The first term in Eq.~\eqref{noneq_chihara} represents the contribution of the \lq{free\rq} electrons to the total structure. The second term (for which there is no analog in the equilibrium Chihara formula) effectively gives a correction (mixing) term for the ionic contribution to dynamic screening at high frequencies. As such, it is small everywhere except near the ion acoustic frequency and vanishes once the Born-Oppenheimer approximation is applied, as expected. The third term is the usual ionic structure contribution convolved with the density of all the electrons (both bound and free) associated with the ions.

The contributions of the \lq{free\rq} electrons to the total DSF (\ref{noneq_chihara}) occur in both the high-frequency term $S_{ee}^{\dagger}$, e.g. via the plasmon resonances, and in the generalized screening cloud $\rho$. In the latter, the free electrons assume the role of a dynamic version of the pure static screening cloud (usually denoted $q_{a}(k)$). Of course, the bound electrons also influence the structure of both terms and cannot formally be separated in the physical picture. However, since Raman transitions are cut off at the binding energy, which is usually significantly larger than the energies characterizing the ion dynamics, then the effect of bound-free transitions may be expected to be naturally strongly localized in the high-frequency term $S_{ee}^{\dagger}$. Conversely, the bound-bound transitions (including the elastic Rayleigh scattering) do not support high-frequency collective excitations and are therefore expected to be localized in the generalized screening cloud, which extends over a similar dynamic range to the ion density response. The bound electron contribution to the generalized screening cloud $\rho$ is expected to behave as a dynamic (complex) version of the ionic form factor contribution to the Rayleigh amplitude in the chemical picture. Thus, the same conceptual separation of free- and bound-electron terms introduced by the chemical picture may be possible to envisage in the physical picture, despite being impractical to realize.

The Chihara-like decomposition for the retarded and advanced quantities related to Eq.~\eqref{neq_chihara_ansatz} reads
\begin{align}
  L_{ee}^{\retadv}
  = 
  {L_{ee}^{\dagger}}^{\!\!\retadv}+\rho^{\retadv} L_{ii}^{\retadv} \rho^{\retadv}
  \,.
\end{align}
From the known expressions for the generalized screening cloud and the ion-ion and electron-ion structure, the retarded/advanced structure of the first term can be obtained to be
\begin{widetext}
\begin{align}
  \label{Lee_dagger_R/A}
  {L_{ee}^{\dagger}}^{\!\!\retadv}
  = &\,
  \frac{1}
  {\qcal_{ii}^{\retadv} (1-\tcal_{ee}^{\retadv})^{2}}
  \left\{
  \qcal_{ee}^{\retadv}\qcal_{ii}^{\retadv} (1-\tcal_{ee}^{\retadv})
  \right.
  \left. - (\qcal_{ei}^{\retadv})^{2} (1-\tcal_{ii}^{\retadv})
  \right\}
  \,.
\end{align}
\end{widetext}
For the case of the DPA, Eq.~\eqref{Lee_dagger_R/A} reduces to
\begin{align}
  \label{free_electron_gas}
  {L_{ee}^{\dagger}}^{\!\!\retadv}
  =
  \lcal_{ee}^{\retadv}
  \,,
\end{align}
as expected. Since electron-ion terms in the polarization functions are neglected in Eq. \eqref{free_electron_gas}, it may now be considered to be a true free-electron gas contribution. This strong simplification means that in many systems, for instance when the free electrons are highly degenerate, for which the electron-ion interaction is indeed weak, the approximation of a free-electron gas structure is valid. However, caution is advised for a general warm dense matter state to always use this approximation as electron-ion correlations might influence the high-frequency electron feature.

The calculation of the correlation function for the high-frequency density response is rather lengthy but straightforward and one obtains
\begin{widetext}
\begin{align}
\label{Lee_dagger_>}
  {L_{ee}^{\dagger}}^{\!\!>}
  = 
  \frac{1}{\left|1-\tcal_{ee}^{\ret}\right|^4
  \left|\qcal_{ii}^{\ret}\right|^2}
  &
  \Big\{\left|1-\tcal_{ee}^{\ret}\right|^2\left|\qcal_{ii}^{\ret}\right|^2
  \left[\qcal_{ee}^{>}\left(1-\tcal_{ee}^{\adv}\right)+\qcal_{ee}^{\adv}\tcal_{ee}^{>}\right]
  \vphantom{\frac{1}{1}}
  -\qcal_{ei}^{>}\qcal_{ei}^{\adv}\left(1-\tcal_{ii}^{\adv}\right)\qcal_{ii}^{\ret}\left|1-\tcal_{ee}^{\ret}\right|^2
  \nonumber\\
  &
  -\qcal_{ei}^{>}\qcal_{ei}^{\ret}\left(1-\tcal_{ii}^{\adv}\right)\qcal_{ii}^{\ret}\left(1-\tcal_{ee}^{\adv}\right)^2
  \vphantom{\frac{1}{1}}
  -{\qcal_{ei}^{\adv}}^2\left(1-\tcal_{ii}^{\adv}\right)\qcal_{ii}^{\ret}\tcal_{ee}^{>}\left(1-\tcal_{ee}^{\ret}\right)
  \nonumber\\
  &
  -\left|\qcal_{ei}^{\ret}\right|^2\left(1-\tcal_{ii}^{\adv}\right)\qcal_{ii}^{\ret}\tcal_{ee}^{>}\left(1-\tcal_{ee}^{\adv}\right)
  \vphantom{\frac{1}{1}}
  +{\qcal_{ei}^{\ret}}^2\left(1-\tcal_{ii}^{\adv}\right)\qcal_{ii}^{>}\left(1-\tcal_{ee}^{\adv}\right)^2
  \nonumber\\
  &
  +{\qcal_{ei}^{\ret}}^2\tcal_{ii}^{>}\qcal_{ii}^{\adv}\left(1-\tcal_{ee}^{\adv}\right)^2
  \Big\}
  \,.
\end{align}
\end{widetext}
Once again, this greatly simplifies in DPA to
\begin{align}
  {L_{ee}^{\dagger}}^{\!\!>}
  =
  \lcal_{ee}^{>}
  \,,
\end{align}
and is again a true free-electron gas contribution. The degree of complexity underpinning the structure of the high-frequency behavior found in \eqa{Lee_dagger_R/A}{Lee_dagger_>} is a reflection of the fact that the cross species contributions to the polarization functions allow for a vastly wider range of routes for density excitations to couple. It is therefore worth stating that a great deal of interesting and complicated physics may have been neglected in the analysis of x-ray scattering data when electron-ion correlations were neglected. In fact, Fortmann et al. pioneered and successfully applied an extended Mermin approach to the dynamic structure of the \lq{free\rq} electron gas in equilibrium taking into account electron-electron as well as electron-ion correlations \cite{FWR_2010}.

\begin{figure*}[t]
  \includegraphics[width=0.9\textwidth,clip=true]{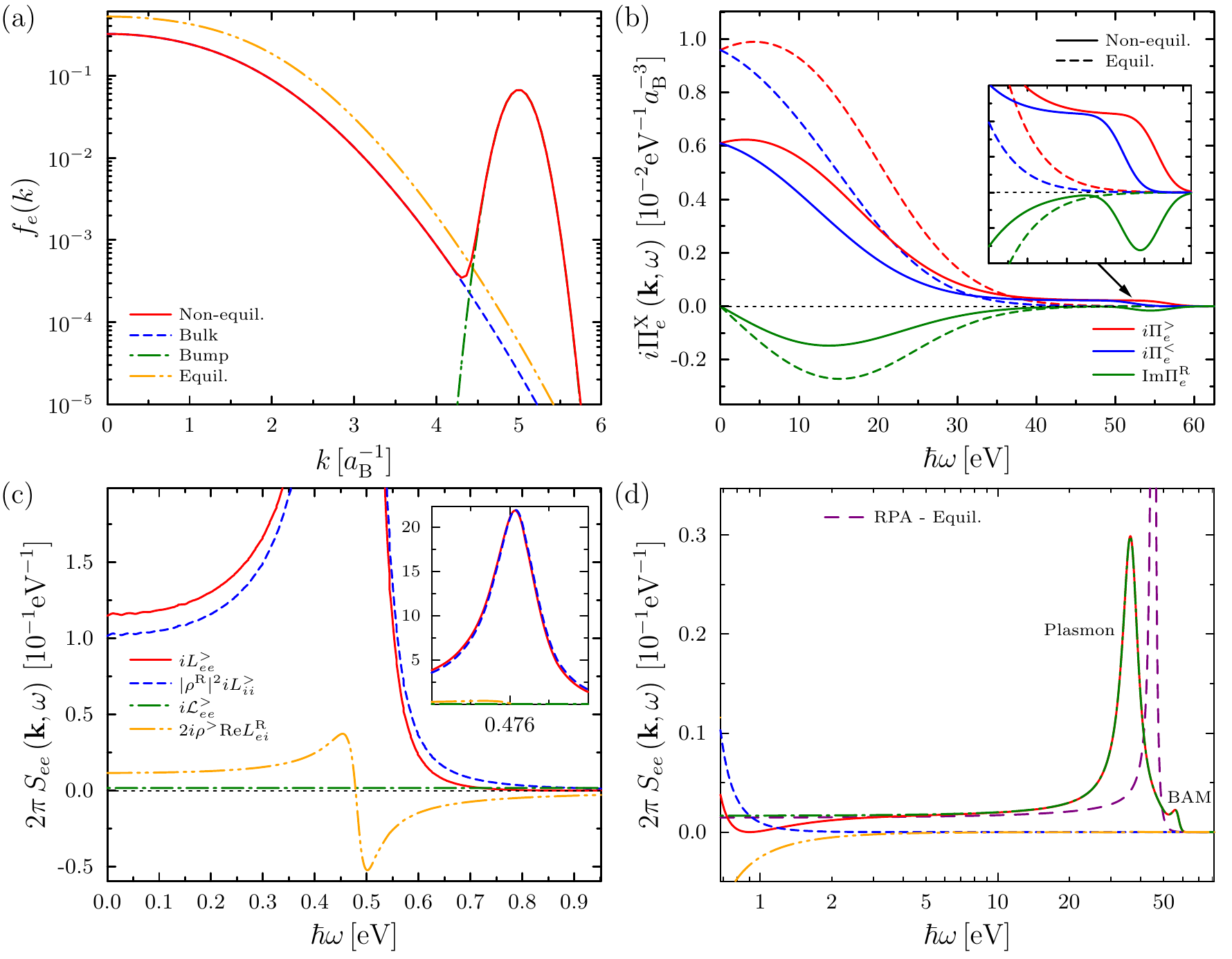}
  \caption{
    Total electron structure factor in RPA for a non-equilibrium dense hydrogen plasma and its Chihara decomposition for $n_{e} = n_{p} = 10^{24} \, \unit{cm^{-3}}$ and $k = 0.4\,a_{\text{B}}^{-1}$. (a): Distribution functions for the electrons. As used in the previous examples, the electron distribution is given by the sum of a bulk component ($T_{c} = 4\times 10^{5} \, \unit{K}$) and a shifted Gaussian bump ($T_{b} = 10^{4} \, \unit{K}$ and $p_{b} = 5 \,\hbar a_{\text{B}}^{-1}$). The protons are in equilibrium at $T_{p} = 10^{5} \, \unit{K}$. In comparison, the equilibrium electron distribution function for these conditions is shown by the orange dash-dotted curve. (b): Correlation functions and the imaginary part of the retarded polarization function. (c): Proton dominated part of the electron structure factor. The contributions from the free-electrons (dot-dashed green curve), the proton structure convoluted with the screening cloud (dashed blue curve) and the electron-ion mix term (double dot-dashed orange curve) are all shown. The resulting total DSF is given by the solid red curve. The inset in panel (c) shows the ion acoustic peak in detail. (d): Electron dominated part of the total electron structure factor (plotted on a logarithmic frequency scale). The labeling of the curves is the same as in panel (c). For comparison, the DSF for the equilibrium system at the same density is shown by the long-dashed purple curve.
    }
      \label{see_ht}
\end{figure*}

\begin{figure*}[t]
  \includegraphics[width=0.9\textwidth,clip=true]{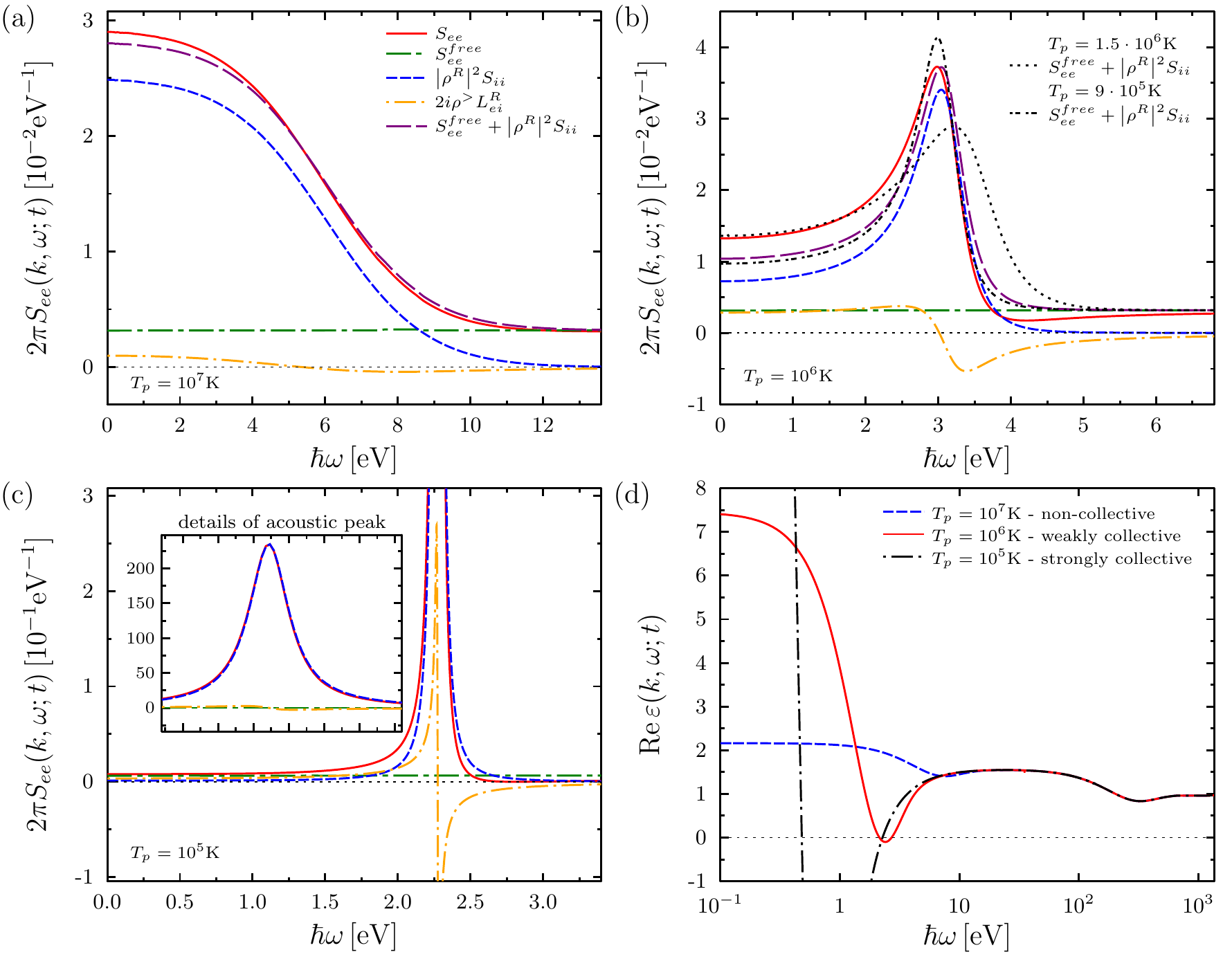}
  \caption{
    Proton acoustic mode part of the total electron structure factor in RPA in two-temperature dense hydrogen and its Chihara decomposition for a total electron density of $n_e=10^{25}\,\unit{cm^{-3}}$ and a wave number of $k = 1 \, a_{\text{B}}^{-1}$. The electron temperature is fixed at $T_{e} = 10^7\,\unit{K}$ in all the examples shown. The proton temperature changes from $T_p = 10^7\,\unit{K}$ (a), to $T_p = 10^6\,\unit{K}$ (b), and $T_p=10^5\,\unit{K}$ (c)).
The black dashed and dotted curves in panel (b) show results one might obtain when trying to fit the red full signal using the Born-Oppenheimer approximation. The inset in panel (c) shows the ion peak in more detail. Panel (d) contains the real part of the two-component (proton and electron) dielectric function for the three different proton temperatures demonstrating the appearance of zeros (ion acoustic modes) at low frequencies.
    }
  \label{see_twotemp}
\end{figure*}

\vspace{1cm}
\subsection{Example for the Chihara decomposition in non-equilibrium}

As an example for a two-component material, we consider hydrogen at a total proton density of $n_p=10^{24} \, \unit{cm^{-3}}$, as shown in panel (a) in Fig. \ref{see_ht}. The ions (protons) are in equilibrium at a temperature of $T=10^5\,\unit{K}$. The electrons are described by a distribution function composed of the sum of a bulk and a Gaussian bump as described in the figure caption. As we want to qualitatively demonstrate the important features of the non-equilibrium Chihara decomposition of the total electron structure, we restrict us to the RPA approximation in the polarization function and subsequent quantities.

Panel (b) of Fig. \ref{see_ht} shows the differences in the polarization correlation functions between equilibrium and non-equilibrium. The magnitude of the non-equilibrium functions is smaller because of the reduced bulk density. The extension in frequency space is comparable with the equilibrium case due to the high energy bump in the distribution function. Panels (c) and (d) of Fig. \ref{see_ht} display the ion acoustic mode (c) and the electron part (d) of the total electron DSF. The ion acoustic mode is exceptionally well developed due to the low ion temperature and the corresponding changes in the electronic screening of the ion-ion interactions. It can further be seen, that the full ion acoustic peak is dominated by the total ion structure factor $i\hbar L_{ii}^{>}$ multiplied by the non-equilibrium electron screening cloud $\left|\rho^{\ret}\right|^2$. Nevertheless, the non-equilibrium free electron gas structure $i\lcal_{ee}^{>}$ and {\em especially} the electron-ion mix term $i2 \rho^{>}\mbox{Re}L_{ei}^{\ret}$ give important contributions. This is a special attribute which appears only for non-equilibrium distribution functions as the latter term vanishes in equilibrium due to the scale separation of electrons and ions which, again only in equilibrium, gives rise to the application of the Born-Oppenheimer approximation \cite{MRWF_2017}. The electron part of the total electron structure factor is given to high accuracy by the free electron structure factor $i\lcal_{ee}^{>}$. The deviation from equilibrium is visible in the reduced height of the plasmon feature at $\omega\sim 2.6 \, \unit{ryd}$ and in the broad feature with a second maximum to the right of the plasmon peak. 

\subsection{Chihara decomposition in a two-temperature systems}

During the relaxation process after energy intake due to a laser or particle beam, or due to a shock wave, there is usually a time span during which the non-equilibrium system may be modeled as a two-temperature system with a temperature ascribed to the electron subsystem and a different temperature to the ion subsystem. Such states have been found in experiments to last for up to several hundred picoseconds and the general agreement in the description of the energy transfer and relaxation process between theory and experiment is not at an acceptable level currently \cite{CNXF_1992,NCXF_1995,WVBC_2012}. XRTS has emerged as a possible diagnostic for such relaxation, shedding light on the time scales and mechanisms \cite{WVBC_2012}. The underlying theory for the electronic structure naturally needs to capture the non-equilibrium physics adequately. Figure \ref{see_twotemp} demonstrates that the current formalism of this paper is well suited to describe two-temperature systems and that a description using an equilibrium structure theory will introduce errors. 

We choose extreme conditions that might occur during inertial fusion. In equilibrium (Fig.~\ref{see_twotemp}, panel (a)), there is no problem with the decomposition of the total electron structure factor into free electron part and ion part. With a reduction of the ion temperature in relation to the electron temperature and under proper coupling of the ion density modes and the electron screening, a screened ion acoustic mode forms (Fig. \ref{see_twotemp}, panel (d), zeros of the real part of the dielectric function, black curve). Once this is the case, the electron-ion cross term in the Chihara formula, comprised of the correlation function of the electron screening cloud and the electron-ion structure, does not vanish anymore and is of the same order as the free electron feature in the spectral range of the ion acoustic mode. This constitutes a breakdown of the Born-Oppenheimer approximation, on which the usual equilibrium Chihara decomposition is founded. For very large temperature differences however, it seems to be the case that the {\em relative} error introduced by neglecting the $i2\rho^{>}\mbox{Re}L_{ei}^{\ret}$ term is tolerable. However, as large temperature differences between species often occur at the beginning of the relaxation process where it is most likely that either one or even both species have not yet fully established their own temperature, it seems prudent to always use the full non-equilibrium formalism.

In panel (b) of Fig. \ref{see_twotemp}, we also try to fit the artificial signal (red curve) using the Chihara formula in Born-Oppenheimer approximation, i.e., without the electron-ion cross term in the second line of Eq.~(\ref{noneq_chihara}). As can be observed, such procedure does not give a good fit of the low frequency behaviour (purple dashed curve). One could try and fit the red curve with different ion temperatures but a full fit of the total ion acoustic signal would not be possible.


\section{Summary}
This paper presents a comprehensive quantum theory for calculating the dynamic properties of correlated, two-component charged-particle systems in non-equilibrium states. Of principal concern to the results is the description of the dynamic structure factor (DSF), which has been presented in a general way that incorporates all levels of inter-particle interactions. The central role played by the polarization functions and, more fundamentally, by the Wigner distribution functions of the electrons and ions was demonstrated. A highly generalized framework has nevertheless been provided, which formalises the correct approach to modeling such systems when better approximations for the polarization functions can be developed. The specific case of the diagonalized polarization approximation (DPA), which arises from neglecting direct electron-ion terms in the polarization function and results in an effective two-fluid description in linear-response, has been discussed. We showed that the DPA description leads to results that agree with previously published results for various quantities of interest, such as the DSF and the energy transfer rate.

In considering the polarization function for correlated systems, we appealed to a perturbation expansion of the fully self-consistent Bethe-Salpeter equation. The resulting expression goes beyond the weakly coupled limit of the random phase approximation (RPA) by including terms of first-order in the interaction potential corresponding to vertex and self energy corrections. We have demonstrated the  evaluation of these terms for non-equilibrium states. Comparison of the vertex and self energy terms with known results of equilibrium limiting cases from the literature shows good agreement in the non-degenerate limit. In the highly degenerate case, the vertex term was shown to give excellent agreement with previous work. However, the self energy term calculated in this work gives qualitatively different results. 

For specific results for the dynamic structure factor, the (potentially time-dependent) non-equilibrium distribution functions of electrons and ions are needed as the fundamental inputs to the theory. In equilibrium, the difference between screened and unscreened polarization function corrections has been demonstrated. It was shown that screening plays an important role in the cancellation of large discrepancies from RPA in Coulomb systems. In general, the exchange and correlation contributions were shown to produce the expected broadening and down-shifting of the quasi-particle excitations. Moreover, our (statically screened) results showed excellent agreement with well-known local field correction schemes. Of course, such agreement is not expected in general, especially at higher coupling strengths.

A simple analytic form for the distribution function as expected in laser-driven experiments has been used to demonstrate non-equilibrium effects. Our results show that the plasmon position, width and amplitude can be significantly affected by large numbers of non-thermal electrons. In particular, we showed exactly how the principle of detailed balance is violated under non-equilibrium conditions. For the specific example considered, a stiffening of the plasmon resonance was observed, suggesting that the effect of exchange and correlation cannot be predicted a priori in non-equilibrium. The dispersion relation shows the formation of the beam acoustic mode, in line with previous studies of similar non-equilibrium systems.

A generalized non-equilibrium electronic screening cloud, i.e. a non-equilibrium form factor, has been introduced. This enables a Chihara-like decomposition of the total electron structure into a free electron part and a part describing the electrons associated with the ion structure, be it as bound electrons or as screening cloud. Such decomposition cannot and does not rely on the Born-Oppenheimer approximation in non-equilibrium and additional terms to the equilibrium decomposition arise due to this feature. Such non-equilibrium decomposition will be of use for experiments trying to create and probe matter on femto-second time scales using ultra-short lasers, XFELs, and XRTS.

The example for laser-driven electrons in warm dense hydrogen shows the non-equilibrium coupling of the electrons to the ions. This leads to an asymmetric change in the peak of the ion acoustic mode. The electron-ion cross term is of similar order as the free electron feature for ion excitation frequencies. The free-electron part of the total electron structure is given to high accuracy by the free electron structure. For two-temperature systems, it was shown that an equilibrium treatment using two different temperatures should be used only for very small temperature differences. For most situations, electron-ion cross terms play a role that cannot be neglected.

The current work offers many possibilities and challenges for future work. For example, it demonstrates the formation of additional channels for electron-ion energy exchange, which may have a significant impact on predictions of temperature relaxation in dense, partially-ionized plasmas. The coupling of micro- and macro-variables needs to be studied in order to take into account gradients in a rigorous way. Most importantly, however, the electron-ion polarization function needs to be treated rigorously, i.e., the electron-ion problem in non-equilibrium needs to be solved in general, since this was shown to be necessary to properly incorporate processes such as bound-free transitions and electron-ion scattering in non-equilibrium systems.


\begin{acknowledgments}
  The authors gratefully acknowledge fruitful discussions with and comments from D.O. Gericke (University of Warwick). J.V. also acknowledges generous hospitality at SIMES, SLAC, Stanford (USA).
\end{acknowledgments}



\section*{Appendix A: Fully-interacting electron-ion density correlation function}
\label{two_comp_derivation_appendix}

In the main text, the system of coupled equations for the density response functions $L_{ab}$ \eqref{L_ab_def} have been written in terms of two sets of auxiliary response functions $\lcal_{ab}$ \eqref{lcal_ab_def} and $\rcal_{ab}$ \eqref{rcal_ab_def}, all of which are defined on the Keldysh time contour. Using the compact notation previously described (in which the coordinates and integrations are supressed) the real-time functions required to evaluate the dynamic structure factor are given by application of the Langreth-Wilkins \cite{LW_1972} rules to the system of equations \eqref{L_ab_def}. Specifically, one finds
\begin{align}
  \label{app_L_ab_><}
  L_{ee}^{\gtrless}
  = &\,
  \lcal_{ee}^{\gtrless} + \rcal_{ee}^{\gtrless}
  + \big( \rcal_{ee}^{\gtrless}V_{ee} + \rcal_{ei}^{\gtrless}V_{ie} \big) L_{ee}^{\adv}
  \nonumber\\ &\,
  + \big( \rcal_{ee}^{\ret}V_{ee} + \rcal_{ei}^{\ret}V_{ie} \big) L_{ee}^{\gtrless}
  \,,
  \nonumber\\
  L_{ei}^{\gtrless}
  = &\,
  \lcal_{ei}^{\gtrless} + \rcal_{ei}^{\gtrless}
  + \big( \rcal_{ei}^{\gtrless}V_{ie} + \rcal_{ee}^{\gtrless}V_{ee} \big) L_{ei}^{\adv}
  \nonumber\\ &\,
  + \big( \rcal_{ee}^{\ret}V_{ee} + \rcal_{ei}^{\ret}V_{ie} \big) L_{ei}^{\gtrless}
  \,,
\end{align}
for the greater/lesser correlation functions and
\begin{align}
  \label{app_L_ab_R/A}
  L_{ee}^{\retadv}
  = &\,
  \lcal_{ee}^{\retadv} + \rcal_{ee}^{\retadv}
  + \big( \rcal_{ee}^{\retadv}V_{ee} + \rcal_{ei}^{\retadv}V_{ie} \big) L_{ee}^{\retadv}
  \,,
  \nonumber\\
  L_{ei}^{\retadv}
  = &\,
  \lcal_{ei}^{\retadv} + \rcal_{ei}^{\retadv}
  + \big( \rcal_{ei}^{\retadv}V_{ie} + \rcal_{ee}^{\retadv}V_{ee} \big) L_{ei}^{\retadv}
  \,, 
\end{align}
for the retarded/advanced functions. The detailed structures of the equations including the coordinates and integrations are identical in form to \eq{L_ee_>_def}. The equivalent correlation and retarded/advanced versions of the auxiliary functions \eqref{lcal_ab_def} are
\begin{align}
  \label{app_curlyL_ab_><}
  \lcal_{ee}^{\gtrless}
  = &\,
  \Pi_{ee}^{\gtrless}
  + \big( \Pi_{ee}^{\gtrless} V_{ee} + \Pi_{ei}^{\gtrless} V_{ie} \big) \lcal_{ee}^{\adv}
  \nonumber\\ &\,
  + \big( \Pi_{ee}^{\ret} V_{ee} + \Pi_{ei}^{\ret} V_{ie} \big) \lcal_{ee}^{\gtrless}
  \,,
  \nonumber\\
  \lcal_{ei}^{\gtrless}
  = &\,
  \Pi_{ei}^{\gtrless}
  + \big( \Pi_{ee}^{\gtrless} V_{ee} + \Pi_{ei}^{\gtrless} V_{ie} \big) \lcal_{ei}^{\adv}
  \nonumber\\ &\,
  + \big( \Pi_{ee}^{\ret} V_{ee} + \Pi_{ei}^{\ret} V_{ie} \big) \lcal_{ei}^{\gtrless}
  \,,
\end{align}
and
\begin{align}
  \label{app_curlyL_ab_R/A}
  \lcal_{ee}^{\retadv}
  = &\,
  \Pi_{ee}^{\retadv}
  + \big( \Pi_{ee}^{\retadv} V_{ee} + \Pi_{ei}^{\retadv} V_{ie} \big) \lcal_{ee}^{\retadv}
  \,,
  \nonumber\\
  \lcal_{ei}^{\retadv}
  = &\,
  \Pi_{ei}^{\retadv}
  + \big( \Pi_{ee}^{\retadv} V_{ee} + \Pi_{ei}^{\retadv} V_{ie} \big) \lcal_{ei}^{\retadv}
  \,,
\end{align}
respectively. Furthermore, for \eqref{rcal_ab_def} we have
\begin{align}
  \label{app_curly_Rab_><}
  \rcal_{ee}^{\gtrless}
  = &\,
  \big( \lcal_{ee}^{\gtrless} V_{ei} + \lcal_{ei}^{\gtrless} V_{ii} \big) \lcal_{ie}^{\adv}
  + \big( \lcal_{ee}^{\ret} V_{ei} + \lcal_{ei}^{\ret} V_{ii} \big) \lcal_{ie}^{\gtrless}
  \,,
  \nonumber\\
  \rcal_{ei}^{\gtrless}
  = &\,
  \big( \lcal_{ee}^{\gtrless} V_{ei} + \lcal_{ei}^{\gtrless} V_{ii} \big) \lcal_{ii}^{\adv}
  + \big( \lcal_{ee}^{\ret} V_{ei} + \lcal_{ei}^{\ret} V_{ii} \big) \lcal_{ii}^{\gtrless}
  \,,
\end{align}
and
\begin{align}
  \label{app_curlyR_ab_R/A}
  \rcal_{ee}^{\retadv}
  = &\,
  \big( \lcal_{ee}^{\retadv} V_{ei} + \lcal_{ei}^{\retadv} V_{ii} \big) \lcal_{ie}^{\retadv}
  \,,
  \nonumber\\
  \rcal_{ei}^{\retadv}
  = &\,
  \big( \lcal_{ee}^{\retadv} V_{ei} + \lcal_{ei}^{\retadv} V_{ii} \big) \lcal_{ii}^{\retadv}
  \,.
\end{align}
For all the equations \eqref{app_L_ab_><}--\eqref{app_curlyR_ab_R/A}, the corresponding ion-ion and ion-electron functions are obtained by swapping labels $e\Leftrightarrow i$ in every term. From these expressions, all the density response properties of a two-component system may be generated. 

In particular the complementary explicit expressions to Eq. (\ref{L_ee_>_result}) for the electron-ion and ion-ion structure factors are
\bea
L_{ei}^>&=&\frac{(1-{\cal T}_{ee}^{\adv})\,{\cal Q}_{ei}^> + {\cal T}_{ee}^>\,{\cal Q}_{ei}^{\adv}}
{|1-{\cal T}_{ee}^{\ret}|^2}\,,\label{leifull}
\\
L_{ii}^>&=&\frac{(1-{\cal T}_{ii}^{\adv})\,{\cal Q}_{ii}^> + {\cal T}_{ii}^>\,{\cal Q}_{ii}^{\adv}}
{|1-{\cal T}_{ii}^{\ret}|^2}\,,\\
L_{ie}^>&=&\frac{(1-{\cal T}_{ii}^{\adv})\,{\cal Q}_{ie}^> + {\cal T}_{ii}^>\,{\cal Q}_{ie}^{\adv}}
{|1-{\cal T}_{ii}^{\ret}|^2}\,.
\eea
Again, all functions depend on a full set of variables $\{\kvec,\omega;t\}$.


\section*{Appendix B: Recovery of results of De Witt et al.}
\label{de_witt_limit_derivation_appendix}

Here, we briefly detail the derivation of the results from De Witt et al. \cite{DSSK_1995} using our more general non-equilibrium framework. The two expressions to consider are the retarded polarization functions for the vertex term \eqref{pirwitt} and self energy \eqref{pisr}. In the non-degenerate case one may use the Maxwell-Boltzmann distribution function
\begin{align}
  \label{app:mb_dist_def}
  f_{a}(k)
  = &\,
  \frac{D_{a}}{s_{a}}\exp\rnd{-\beta\frac{\hbar^{2}k^{2}}{2m_{a}}}
  \,,
\end{align}
where $D_{a} = n_{a}\rnd{2\pi\beta\hbar^{2}/m_{a}}^{3/2}$ is the degeneracy parameter and $s_{a} = 2\sigma_{a}+1$ is the spin summation for fermions. Normalizing the dummy wave vector integrations according to $x^{2} = \beta \hbar^{2} q^{2}/2m_{a}$, $x'^{2} = \beta \hbar^{2} q'^{2}/2m_{a}$ and also defining $y^{2} = \beta\hbar^{2}k^{2}/2m_{a}$ then for $\omega = 0$ one finds
\newcommand{\xvec}{\mathbf{x}}
\newcommand{\yvec}{\mathbf{y}}
\begin{align}
  \label{app:pi_v_00_1}
  {\Pi_{a}^{\text{V}}}^{\text{R}}(\kvec,0)
  = &\,
  -\frac{n_{a}\beta\rnd{\kappa_{\text{D}a}\lambda_{a}}^{2}}{2\pi^{3}s_{a}}
  \int \dif\xvec\,\dif\xvec' \, \frac{e^{-(x^{2} + x'^{2})}}{(\xvec - \xvec')^{2}}
  \nonumber\\
  &\times\frac{e^{-2\xvec\cdot\yvec - y^{2}} - 1}{-2\xvec\cdot\yvec - y^{2}}\,\frac{e^{-2\xvec'\cdot\yvec - y^{2}} - 1}{-2\xvec'\cdot\yvec - y^{2}}
  \,,
\end{align}
for the vertex term and
\begin{align}
  \label{app:pi_s_00_1}
  {\Pi_{a}^{\text{S}}}^{\text{R}}(\kvec,0)
  = &\,
  -\frac{n_{a}\beta\rnd{\kappa_{\text{D}a}\lambda_{a}}^{2}}{2\pi^{3}s_{a}}
  \int \dif\xvec\,\dif\xvec' \, \frac{e^{-(x^{2}+x'^{2})}}{(\xvec - \xvec')^{2}}
  \nonumber\\
  & \times
  \left[
  \frac{e^{-2\xvec\cdot\yvec - y^{2}} - 1}{\rnd{-2\xvec\cdot\yvec - y^{2}}^{2}} - \frac{1}{-2\xvec\cdot\yvec - y^{2}} 
  \right.
  \nonumber\\
  &\left.
  +\frac{1 - e^{2\xvec'\cdot\yvec - y^{2}}}{\rnd{2\xvec'\cdot\yvec - y^{2}}^{2}} - \frac{1}{2\xvec'\cdot\yvec - y^{2}} 
  \right]
  \,,
\end{align}
for the self energy term. In the long-wavelength limit $(y \to 0)$ one may expand the exponentials featuring $y$. In \eq{app:pi_v_00_1} only the first order needs to be retained, whereas for \eq{app:pi_s_00_1} the squared denominators require expansion to second order. All terms featuring $y^{n}$ with $n>2$ can be neglected. The second line of the vertex term immediately gives unity. For the self energy term, one finds
\begin{align}
  \label{app:pi_s_00_2}
  {\Pi_{a}^{\text{S}}}^{\text{R}}(\kvec\to0,0)
  = &\,
  -\frac{n_{a}\beta\rnd{\kappa_{\text{D}a}\lambda_{a}}^{2}}{2\pi^{3}s_{a}}
  \int \dif\xvec\,\dif\xvec' \, \frac{e^{-(x^{2}+x'^{2})}}{(\xvec - \xvec')^{2}}
  \nonumber\\
  & \times
  \left[
  \frac{2(\xvec\cdot\yvec)^{2}}{\rnd{2\xvec\cdot\yvec + y^{2}}^{2}}
  +\frac{2(\xvec'\cdot\yvec)^{2}}{\rnd{2\xvec'\cdot\yvec - y^{2}}^{2}}
  \right]
  \,.
\end{align}
Neglecting the $y^{2}$ terms in the denominators of the second line of \eqref{app:pi_s_00_2} again gives unity. Thus, both terms require evaluation of the following dimensionless integral
\begin{align}
  \label{app:pi_00_integral}
  I
  = &\,
  \int \dif\xvec\,\dif\xvec' \, \frac{e^{-(x^{2}+x'^{2})}}{(\xvec - \xvec')^{2}}
  = \pi^{3}
  \,.
\end{align}
The result given in the main text \eqref{dewitt_result} then immediately follows.




\end{document}